\RequirePackage{ifpdf}
\documentclass[nohyper,12pt,letterpaper]{JHEP3}
\pdfoutput=1
\usepackage{epsfig}
\usepackage[latin1]{inputenc}
\usepackage{bbm,amsfonts}
\usepackage{graphicx}
\usepackage{amssymb,amsmath}
\usepackage{fancybox,framed}
\usepackage{dsfont}
\usepackage{mathtools}
\usepackage{braket}
\usepackage{cite}
\usepackage{slashed}

\author{Marco S. Bianchi\\\\
Centre for Research in String Theory,
School of Physics and Astronomy\\
Queen Mary University of London,
Mile End Road, London E1 4NS, UK \\
\qquad\\
E-mail: \email{ m.s.bianchi@qmul.ac.uk}
}

\abstract{We compute the simplest one-loop planar amplitudes in Higgsed ABJM theory at a generic point of the moduli space. We explicitly check that they can be expressed in terms of integrals which are invariant under dual conformal symmetry involving masses, in a similar fashion as in ${\cal N}=4$ SYM.
}

\preprint{February 2015\\QMUL-PH-15-03}

\title{A note on scattering amplitudes on the moduli space of ABJM}


\csname @addtoreset\endcsname{equation}{section}

\def\bseq{\begin{subequation}}  
\def\eseq{\end{subequation}}
\def\bsea{\begin{subeqnarray}}  
\def\esea{\end{subeqnarray}}

\newcommand{\beq}{\begin{equation}}
\newcommand{\bea}{\begin{eqnarray}}
\newcommand{\eea}{\end{eqnarray}}
\newcommand{\eeq}{\end{equation}}

\def\beq{\begin{equation}}
\def\eeq{\end{equation}}
\def\bea{\begin{eqnarray}}
\def\eea{\end{eqnarray}}
\def\Tr{\mathrm{Tr}}

\begin{document}

\allowdisplaybreaks

\section{Introduction}

In the context of ${\cal N}=4$ SYM a recipe for computing amplitudes at strong coupling via the AdS/CFT correspondence was spelled out in \cite{Alday:2007hr}.
Such a description makes manifest certain symmetry properties that amplitudes display in their perturbative expansion at weak coupling.
In particular dual conformal symmetry is naturally mapped to the standard conformal invariance of Wilson loops through the amplitude/Wilson loop duality \cite{Drummond:2007cf,Drummond:2007au,Brandhuber:2007yx}. 
For superamplitudes dual superconformal and Yangian symmetry \cite{Drummond:2006rz,Drummond:2007aua,Drummond:2008vq,Brandhuber:2008pf,Drummond:2009fd}  of planar amplitudes is interpreted at strong coupling as the invariance of the $AdS_5\times S^5$ $\sigma$-model under fermionic T-duality \cite{BM,Beisert:2008iq}.

Elaborating on this argument the authors of \cite{Alday:2009zm} claimed that an extension of dual conformal symmetry involving masses also holds for amplitudes away from the origin of the moduli space, whose study was pioneered in \cite{Schabinger:2008ah}. More precisely, interpreting masses as an additional component of dual variables, ordinary dual conformal symmetry naturally extends to invariance under inversions in one extra dimension.
Amplitudes of particles acquiring mass via the Higgs mechanism obey such a symmetry, which on the one hand is a powerful constraint for the integrals appearing in their loop corrections and on the other hand drastically simplifies the computation of the relevant integrals themselves.
In particular, such a picture was suggested to provide a natural and symmetry preserving way of regularizing the infrared divergences of planar amplitudes. Namely, the mass of particles running in the outermost propagators of planar loop integrals is used as a regulator. In order to do this, one specializes to a configuration where all such masses are equal (and external particles are massless) and then takes the small mass limit, keeping only leading order terms in such an expansion.
It has been checked to three and four loops that this regularization is such that amplitudes display a BDS-like \cite{BDS} form \cite{Henn:2010bk,Henn:2010ir}.

Another interesting configuration is four-point scattering with two different masses, and the limit where one is much larger than the other. Then the small mass serves as a regulator of soft infrared singularities in a Bhabha scattering process of two heavy W-bosons. Interestingly, from the coefficient of such a divergence one can extract the loop corrections to the anomalous dimension $\Gamma_{1/2}(\phi)$ of a space-like cusp (at an angle $\phi$ related to the kinematics of the scattering event) between two $1/2$-BPS rays \cite{Henn:2010bk,Correa:2012nk}. This and the high precision at which ${\cal N}=4$ SYM scattering processes are known constitutes a powerful way of computing $\Gamma_{1/2}(\phi)$. Moreover, from the space-like 1/2-BPS cusp one can extract the first perturbative coefficients of the Bremsstrahlung function, which were used to test the formula determined in \cite{Correa:2012at} for its exact value.
As a further development, the picture described above was also applied to the study of bound states of W-bosons, tightened together by the exchange of massless particles associated to the unbroken gauge symmetry.
In particular the dual conformal symmetry exhibited by amplitudes in partially Higgsed ${\cal N}=4$ SYM is pivotal in the computation of the spectrum of W-bosons bound states as shown in \cite{Caron-Huot:2014gia}.

Since the idea of considering amplitudes on the moduli space of ${\cal N}=4$ SYM has triggered such interesting advances, it is a natural question to try to investigate this in other theories.
In this note I consider my favourite one, namely ABJM, and move the first steps towards understanding how much of the ${\cal N}=4$ machinery can be applied to this three-dimensional CFT.

In ABJM theory \cite{ABJM} a strong coupling motivation for amplitudes to respect dual conformal symmetry has not been uncovered. In particular, despite several attempts \cite{Adam:2009kt,Adam:2010hh,DO,Bakhmatov:2010fp,Bakhmatov:2011aa,OColgain:2012si}, a recipe for a fermionic T-duality leaving invariant the corresponding $AdS_4$ $\sigma$-model has not been determined.
Yet, the available results for amplitudes at weak coupling hint that dual conformal and Yangian \cite{Drummond:2009fd} symmetry play a crucial role in scattering processes in ABJM, at least perturbatively.
In particular Yangian \cite{BLM} and dual superconformal \cite{HL2} symmetry of ABJM tree level amplitudes was pointed out and the computation of their loop corrections at one \cite{Bianchi:2012cq,Bargheer:2012cp,Brandhuber:2012un,Brandhuber:2012wy}, two \cite{CH,BLMPS1,BLMPS2,CaronHuot:2012hr} and three \cite{Bianchi:2014iia} loops reveals that they can be expressed in terms of dual conformally invariant integrals.
This and the fact that planar ABJM theory possesses similar signs of integrability as for ${\cal N}=4$ SYM suggest that its on-shell sector could be integrable, despite the lack of strong coupling arguments.

But what happens if we move around in the moduli space of ABJM?
Such a question has been first addressed in \cite{CaronHuot:2012hr}.
The authors analysed the spectrum of masses arising from giving one of the scalar fields of the theory a vacuum expectation value, finding a remarkable resemblance with respect to ${\cal N}=4$ SYM. This motivated the authors to use Higgsing as a regulator for amplitudes in a similar manner as proposed in \cite{Alday:2009zm}. In particular, in order to do this one sets up a configuration such that all massive particles running in loop diagrams have equal mass $m$ and that external particles are massless. Then the limit $m\rightarrow 0$ is taken, keeping ${\cal O}(m^0)$ terms, which provides the Higgs regularized result for the amplitude.
In order to do this practically an effective prescription was taken in \cite{CaronHuot:2012hr}, which seems to lead to a very similar dictionary between logarithms of masses and poles in the dimensional regularization parameter as in ${\cal N}=4$ SYM. This is remarkable as there is no a priori guarantee that infared divergences and their different regularizations should behave in the same manner in different dimensions. In particular the coefficient of the cusp anomalous dimension of the amplitude coincides with that of dimensional regularization.
Apart from their application as a regularization procedure, I find the symmetries of ABJM amplitudes in a nontrivial vacuum interesting in their own respect.
As remarked in \cite{CaronHuot:2012hr} the hints at integrability in ABJM scattering suggest that the symmetry properties exhibited by amplitudes at the origin of the moduli space could carry over also away from it.
On the other hand I think that the absence of a sound argument at strong coupling motivates testing this optimistic expectation against some healthy perturbative computation.
It is the scope of this letter to provide such an explicit check.

The main prediction we want to verify here concerns the symmetry properties of loop integrands.
In \cite{Alday:2009zm} it was claimed that loop integrands appearing in perturbative corrections to amplitudes are invariant under a particular extension of dual conformal symmetry involving masses.
This was tested successfully against the computation of a sufficiently simple one-loop four-point amplitude of scalars.
In this letter we perform an analogous test, namely we compute a simple scalar amplitude and check the symmetry properties of the integrand under extra-dimensional inversions.
In order to do this we do not assume anything and perform a direct computation with Feynman diagrams, keeping all contributions, including bubbles and tadpoles. Indeed, while it is fair to exclude them a priori in four dimensions, as they would contribute with UV divergent integrals, this is not the case in three.
This requires computing the full Lagrangian of Higgsed ABJM theory, since we have not found such a computation carried out completely in literature. This is done in section \ref{sec:Lagrangian}.
Equipped with such a Lagrangian we then derive the relevant Feynman rules required for our computation.

Inspection of the propagators and vertices of the theory selects the easiest amplitudes to compute.
Keeping only scalar fields as external particles such a simple amplitude is arguably the six-point one, with a suitable choice of flavour indices, so as to minimize the number of contributing diagrams.
In section \ref{sec:amplitude} we compute such an amplitude at one loop and verify that indeed its integrand is invariant under the desired extended dual conformal symmetry.
In particular the denominators of the relevant triangle integrals all get masses, curing possible infrared singularities (which are nonetheless invisible in dimensional regularization). They are of the same form as those appearing in computations within ${\cal N}=4$ SYM, namely obtained by replacing squared invariants of dual coordinates with those of the extra-dimensional points, endowed with a mass.
Since the same amplitude vanishes at the origin of the moduli space, one expects it to be proportional to powers of the masses in the numerator, trivializing the small mass limit.
This is indeed the case.
In fact the powers and labels of these masses are exactly such that the required good properties under "four-dimensional" inversion are indeed satisfied.
The possible appearance of these numerators was pointed out in \cite{Henn:2010bk}, where nevertheless the authors argued that they would not affect the BDS exponentiation properties of three-loop Higgs regularised amplitudes of ${\cal N}=4$ SYM which were under exam there.

As a byproduct we also compute an even simpler amplitude, namely a totally fermionic six-point one, which also happens to receive a very limited amount of quantum corrections.
Again, for this to occur flavours have to be selected wisely.
In this case we find the emergence of integrals with the same massive denominators as before. The result then is not directly invariant under dual conformal transformations (though it was so in the massless case), due to the polarization spinors of fermions, which do not transform covariantly.
In fact, in order to really ascertain the symmetry properties of this amplitude one would have to construct the proper superamplitude and dual superconformal generators. Still, we stress that no lower topologies than triangle integrals appear (which is the basic requirement from dual conformal symmetry in three dimensions) and that the denominators are precisely the fourth-dimensional extension of those in the massless case.
Interestingly, also the numerator of this integral looks exactly like a natural extension of the result in the massless case. In particular, elaborating on the extra-dimensional interpretation of masses, the  numerator can be obtained by replacing three-dimensional polarization spinors for massless fermions with four-dimensional ones, which are of the same form of massive ones in three dimensions, provided an identification between the mass and the extra momentum component is made.

\section{Higgsed theory amplitudes}\label{sec:review}

In this section we briefly review the setting of \cite{Alday:2009zm} for amplitudes on the moduli space of ${\cal N}=4$ SYM and its extension to the ABJM case.
${\cal N}=4$ SYM with gauge group $U(N+M)$ is the low energy theory living on a stack of $(N+M)$ $D3$ branes. One can engineer spontaneous symmetry breaking by pulling $M$ branes apart from the other $N$. This would lead to a breaking of the original gauge symmetry to $U(N)\times U(M)$. Further displacing the $M$ branes among themselves one breaks the symmetry to $U(N)\times U(1)^M$. Supposing for simplicity that the branes are moved in only one of the transverse directions, say the 9th, then this would correspond to equipping the adjoint scalars $X^9$ with an expectation value (which is a diagonal matrix with $M$ nonzero entries).
Strings connecting the bunch of $N$ D3's with the separated ones give rise to "heavy" massive particles, such as the W-bosons, whereas excitations of strings stretching between a pair of the $M$ separated branes represent "light" massive particles.

In \cite{Alday:2009zm} the planar scattering of light particles was considered. At loop level one can conveniently take the large $N$ limit, more precisely $N\gg M$, which selects diagrams with the leading number of loops of indices in the unbroken part of the gauge group.
In the planar limit only diagrams survive which have heavy particles running in the outermost propagators, where the external particles attach.
It was then argued that amplitudes constructed in this way enjoy invariance under an extension of dual conformal symmetry involving masses.
This was given an extra-dimensional interpretation, endowing dual coordinates with an additional component representing a mass
\begin{equation}\label{eq:extension}
x_i \rightarrow \hat x_i \equiv (x_i,m_i)
\end{equation}
Then integrands can be rewritten in terms of five-dimensional quantities and are invariant under the five-dimensional inversions
\begin{equation}\label{eq:inversion}
\hat x_i \rightarrow \frac{\hat x_i}{\hat x_i^2}
\end{equation}
for any point, where the understood scalar product is now five-dimensional.
In particular integrands can be expressed in terms of only five-dimensional quantities, provided a $\delta$-functions ensures the integration is four-dimensional. The requirement that the internal point has vanishing conformal weight under inversions excludes bubbles and triangles, as in the massless case.
Then for external points the amplitude is invariant under the extended dual conformal boost
\begin{equation}\label{eq:generator}
K^{\mu} = \sum_i \left[ 2\, x_{i}^{\mu}\, \left( x_{i}^{\nu}\, \frac{\partial}{\partial x_i^\nu} + m_i\, \frac{\partial}{\partial m_i}\right) + \left( x_i^2 + m_i^2 \right) \frac{\partial}{\partial x_{i\,\mu}} \right]
\end{equation}
where $i=1,\dots n$ for $n$-particle scattering and $\mu=0,1,2,3$.
This entails invariance under five-dimensional inversions and dilatations and four-dimensional Poincar\'e group transformations.
Since the integrals constructed in this way are finite, having infrared divergneces regularized by the masses, \eqref{eq:generator} is an exact symmetry of the amplitude also after integration, that is it does not possess an anomaly (as with dimensional regularization).

In ABJM theory spontaneous symmetry breaking was studied in detail in \cite{Berenstein:2008dc,Lee:2010hk}.
In the M-theoretical strong coupling description this is achieved by displacing $M$ M2 branes from a stack of other $N$.
Breaking symmetry in the same fashion as described above yields a spectrum of masses which displays some similarities with the ${\cal N}=4$ SYM one \cite{CaronHuot:2012hr}.
Despite the fact that the strong coupling interpretation of amplitudes and their symmetries is not transparent in ABJM, one could straightforwardly study the fate of dual conformal symmetry away from the origin of the moduli space in the weak coupling perturbative expansion.
To accomplish this task we focus on the same configuration described above, namely we take ABJM with gauge group $U(N+M)\times U(N+M)$ and break it to $U(N)\times U(N)$ plus a bunch of $U(1)$'s.
Then we consider again scattering of light particles in the $N\gg M$ regime, which ensures planarity of the diagrams and a frame of heavy particles running in the outermost propagators.
This setting was already considered in \cite{Alday:2009zm} to motivate Higgs mechanism regularization.
Here we borrow the same construction, while keeping different finite masses and analyse the symmetry properties of amplitudes.
Two things are needed for this: first the complete Lagrangian of Higgsed ABJM and the following Feynman rules, and second a sufficiently simple amplitude to compute.
The first task is carried out in the following section, whereas the choice of the suitable amplitude and its computation are dealt with in section \ref{sec:amplitude}.

\section{Higgsed ABJM Lagrangian}\label{sec:Lagrangian}

In this section we compute the Lagrangian of Higgsed ABJM.
Our starting point is the Lagrangian of \cite{MOS1} for ABJM theory in three-dimensional Minkowski space with signature $(-,+,+)$ (see Appendix \ref{app:A} for more details), before gauge fixing.
\begin{align}
{\cal L}_{ABJM} &=\frac{k}{4\pi}\, \Tr\Big[
A_\alpha
\epsilon^{\mu\rho\nu}\partial_\rho
-\hat A_\mu
\epsilon^{\mu\rho\nu}\partial_\rho
\hat A_\nu
+Y_A^\dagger\partial_\mu\partial^\mu Y^A+i\psi^{\dagger B}\slashed{\partial}\psi_B
\Big]\\
& + \frac{k}{4\pi}\, \Tr\Big[
\frac{2}{3}i\epsilon^{\alpha\beta\gamma}(A_\alpha A_\beta A_\gamma
-\hat A_\alpha\hat A_\beta\hat A_\gamma)\\
&\phantom{=\frac{k}{4\pi}}
-iA_\mu Y^A\overset{\leftrightarrow}{\partial^\mu}Y_A^\dagger
-i\hat A_\mu Y_A^\dagger\overset{\leftrightarrow}{\partial^\mu}Y^A
-\psi^{\dagger B}\slashed{A}\psi_B
+\hat A_\mu\psi^{\dagger B}\gamma^\mu\psi_B \label{eq:cubic}\\
&\phantom{=\frac{k}{4\pi}}
+2Y_A^\dagger A_\mu Y^A\hat A^\mu-\hat A_\mu\hat A_\mu Y_A^\dagger Y^A
-A_\mu A_\mu Y^AY_A^\dagger \label{eq:quartic}\\
&\phantom{=\frac{k}{4\pi}}
+\frac{1}{12}Y^AY_B^\dagger Y^CY_D^\dagger Y^EY_F^\dagger
(\delta_A^B\delta_C^D\delta_E^F
+\delta_A^F\delta_C^B\delta_E^D
-6\delta_A^B\delta_C^F\delta_E^D
+4\delta_A^D\delta_C^F\delta_E^B) \label{eq:scalar}\\
&\phantom{=\frac{k}{4\pi}}
-\frac{i}{2}(
Y_A^\dagger Y^B\psi^{\dagger C}\psi_D
-\psi_D\psi^{\dagger C}Y^BY_A^\dagger)
(\delta_B^A\delta_C^D-2\delta_C^A\delta_B^D)\label{eq:Yukawa2}\\
&\phantom{=\frac{k}{4\pi}}
+\frac{i}{2}\epsilon^{ABCD}Y_A^\dagger\psi_BY_C^\dagger\psi_D
-\frac{i}{2}\epsilon_{ABCD}Y^A\psi^{\dagger B}Y^C\psi^{\dagger D}
\Big] \label{eq:Yukawa1}
\end{align}
where an explanation for all indices can be found at the end of this section.
Without loss of generality we choose to give expectation value to the scalar fields $Y^1$ (see below for an explanation on indices) 
\begin{equation}\label{eq:Higgsing}
(Y^1)^{I}_{\phantom{I}\hat J} \rightarrow v_{i}\, \delta^i_{\phantom{i}\hat j} + (Y^1)^{I}_{\phantom{I}\hat J} \qquad (Y^{\dagger}_{1})^{\hat I}_{\phantom{\hat I}J} \rightarrow \bar v^{i}\, \delta^{\hat i}_{\phantom{\hat i}j} + (Y^{\dagger}_{1})^{\hat I}_{\phantom{\hat I}J}
\end{equation}
meaning that scalar fields acquire vacuum expectation value $v_i$ in the $i=N+1,\dots N+M$ diagonal entries.
This way we break the original $U(N+M)\times U(N+M)$ gauge symmetry to $\left(U(N)\times U(1)^M\right)\times\left(U(N)\times U(1)^M\right)$. The choice of vacuum \eqref{eq:Higgsing} also breaks the original $SU(4)$ flavour symmetry to $SU(3)$, rotating the three remaining scalars with trivial vev.
Then one obtains a new Lagrangian containing extra terms which we collect as follows
\begin{equation}
\hat{\cal L}_{ABJM} = {\cal L}_{ABJM} + {\cal L}_{Higgs}
\end{equation}
In the following subsections we spell out the various contributions to ${\cal L}_{Higgs}$ emerging from the original Lagrangian.

A plethora of indices with different meanings arises.
To avoid confusion we explain our notation as follows. 
We start with gauge indices: we label $I,J,\dots\, = 1,\dots N+M$ gauge indices of the first $U(N+M)$ gauge group, which we split into $a,b,\dots\, = 1,\dots N$ and $i,j,\dots\, = N+1,\dots N+M$, namely the former refer to the unbroken part of the original gauge group, whereas the latter to the broken. We use hatted indices for the second gauge group.
At the price of introducing painfully looking formulae, we spell out all indices for the sake of clarity in what follows.  
We denote with $A,B,\dots\, = 1,2,3,4$ flavour indices for the matter fields and use a hat to distinguish $\hat{A}, \hat{B},\dots\, = 2,3,4$ in the $SU(3)$ subgroup into which the original $SU(4)$ flavour symmetry breaks after the Higgsing \eqref{eq:Higgsing}.
Finally, we reserve Greek letters for spinor indices.

\subsection{Gauge-scalar sector}

Starting from cubic interaction terms between scalar and gauge fields \eqref{eq:cubic} we get additional quadratic pieces
\begin{align}
& i \left[ v_i \left( - (A_{\mu})^I_{\phantom{I}i}\, \delta^i_{\phantom{i}\hat i}\, \partial^{\mu} (Y^\dagger_1)^{\hat i}_{\phantom{\hat i} I} +  
 (\hat A_{\mu})^{\hat i}_{\phantom{\hat i}\hat I}\, \partial^{\mu} (Y^\dagger_1)^{\hat  I}_{\phantom{\hat I} i}\, \delta^i_{\phantom{i} \hat i} \right)
+
\right. \nonumber\\ & \left. +
\bar v^i \left( (A_{\mu})^i_{\phantom{i} I}\, \partial^{\mu} (Y^1)^I_{\phantom{I} \hat i}\, \delta^{\hat i}_{\phantom{\hat i} i} -  
 (\hat A_{\mu})^{\hat I}_{\phantom{\hat I} \hat i}\, \delta^{\hat i}_{\phantom{\hat i} i}\, \partial^{\mu} (Y^1)^i_{\phantom{i} \hat I} \right)
\right]
\end{align}
which would mix gauge and scalar fields.
These unwanted couplings can be cancelled by a proper $R_{\xi}$ gauge
\begin{align}
{\cal L}_{g.f.} &= -\frac{1}{\xi} 
\left( \partial_{\mu} A^{\mu} + i\, \xi\, v_i\, \delta^i_{\phantom{i}I}\, Y^\dagger_1 \right)^I_{\phantom{I}J} \left( \partial_{\mu} A^{\mu} -  i\, \xi\, Y^1\, \bar v^i\, \delta_{i}^{\phantom{i}I} \right)^J_{\phantom{J}I} + \nonumber\\&
+ \frac{1}{\hat \xi}
\left( \partial_{\mu} \hat A^{\mu} - i\, \hat \xi\, v_{\hat j}\, \delta^{\hat j \hat J}\, Y^\dagger_1 \right)^{\hat I}_{\phantom{\hat I} \hat J} \left( \partial_{\mu} \hat A^{\mu} + i\, \hat \xi\, Y^1\, \bar v^{\hat j}\, \delta_{\hat j \hat J} \right)^{\hat J}_{\phantom{\hat J} \hat I}
\end{align}
where the gauge parameters $\xi$ and $\hat \xi$ have dimensions of mass.
The corresponding ghost Lagrangian, although it is not required for the one-loop amplitude we will be interested in, features the standard part
\begin{equation}
{\cal L}_{ghost} = \frac{k}{4\pi}\, \Tr \left[ c^\ast\partial_\mu\partial^\mu c + 
\hat c^\ast\partial_\mu\partial^\mu \hat c
-iA^\mu [c,\partial_\mu c^\ast]
-i\hat A^\mu [\hat c,\partial_\mu\hat c^\ast] \right]
\end{equation}
plus interaction terms between ghosts and $Y^1$ fields arising from the gauge variation of the scalar dependent part of the gauge fixing function, of the form
\begin{equation}
i\, \xi\, \left[ v_i\, (c^\ast)^I_{\phantom{I}i}\, \delta^i_{\phantom{i}\hat i} (Y^{\dagger}_1)^{\hat i}_{\phantom{\hat i}J}\, c^J_{\phantom{J}I} 
- \bar v^i\, (c^\ast)^J_{\phantom{J}I}\, (Y^1)^I_{\phantom{I}\hat i}\, \delta^{\hat i}_{\phantom{\hat i}i}\, c^i_{\phantom{i}J} \right]
\end{equation}
and similarly for ghosts associated to the second gauge group symmetry.
What is left are $(\xi,\hat\xi)$-dependent YM-like kinetic terms for the gauge bosons and a gauge dependent mass for $Y^1$ scalars
\begin{equation}
- \left[ \frac{1}{\xi} (\partial_{\mu} A^{\mu})^2 - \frac{1}{\hat \xi} (\partial_{\mu}\hat A^{\mu})^2 \right]
- |v_i|^2 \left( \xi\, (Y^\dagger_1)^{\hat i}_{\phantom{\hat i}J}\, (Y^1)^J_{\phantom{J}\hat i} - \hat \xi\, (Y^1)^i_{\phantom{i}\hat J}\, (Y^\dagger_1)^{\hat J}_{\phantom{\hat J}i} \right)
\end{equation}
The $Y^1$ fields play the role of Goldstone bosons and, as their gauge dependent mass suggests, they do not correspond to any physical state. In particular, they are not produced in physical processes and it is meaningless to compute amplitudes for them. 
Following with the new terms after Higgsing, from quartic interactions \eqref{eq:quartic} we get new cubic vertices with the $Y^1$ fields and two gauge bosons.
Since, as we have just said, there are no amplitudes $Y^1$ fields, these new vertices do not play any role for computing the amplitudes we are interested in, at first order in perturbation theory. Therefore we do not spell them out here (but in the Appendix \eqref{eq:cubicgaugeY1}).
Finally there are mass terms for the gauge fields
\begin{align}
- |v_i|^2\, (A_{\mu})^i_{\phantom{i}J}\, (A^{\mu})^J_{\phantom{J}i} - |v_i|^2\, (\hat A_{\mu})^{\hat i}_{\phantom{\hat i} \hat J}\, (\hat A^{\mu})^{\hat J}_{\phantom{\hat J}\hat i}
+ 2\, \bar v^i\, v_j\, (A_{\mu})^i_{\phantom{i}j}\, \delta^{j}_{\phantom{j}\hat j}\, (\hat A^{\mu})^{\hat j}_{\phantom{\hat j}\hat i}\, \delta^{\hat i}_{\phantom{\hat i}i}
\end{align}
from which we see that there is a non-trivial mixing between the two gauge groups.
This could be a little annoying when performing computation, although we note that there is no mixing for heavy gauge fields with indices $ia$, namely the W-bosons. According to the discussion in section \ref{sec:review}, we restrict to heavy fields running in the loop, through a large $N$ limit. In this approximation the two gauge fields are not mixing and there is no need to compute a mixed propagator for them (which can be found in \cite{CaronHuot:2012hr}, anyway).

Identifying $|v_i|^2=2\, m_i$ we can derive a very similar formula with respect to the ${\cal N}=4$ case.
Indeed we see that, as already analysed in \cite{Lee:2010hk} and \cite{CaronHuot:2012hr}, the diagonal fields remain massless whereas the off-diagonal heavy ones get masses $m_i^2$ and the light modes have masses $(m_i-m_j)^2$.

\subsection{Yukawa interactions}

From the Yukawa terms of the superpotential \eqref{eq:Yukawa2} we get additional interaction vertices
\begin{align}\label{eq:cubic-Yukawa}
-\frac{i}{2}\, & \left[
v_i\, (Y^\dagger_1)^{\hat J}_{\phantom{\hat J} i}\, \delta^{i}_{\phantom{i}\hat i} (\psi^{\dagger A})^{\hat i}_{\phantom{\hat i} K}\, (\psi_A)^K_{\phantom{K}\hat J} + 
\bar v^i\, (Y^1)^i_{\phantom{i}\hat J}\, (\psi^{\dagger A})^{\hat J}_{\phantom{\hat J}K}\, (\psi_A)^K_{\phantom{K}\hat i}\, \delta^{\hat i}_{\phantom{\hat i}i} +
\right.\nonumber\\&
- \bar v^i\, (\psi_A)^i_{\phantom{i}\hat K}\, (\psi^{\dagger A})^{\hat K}_{\phantom{\hat K} J}\, (Y^1)^J_{\phantom{J}\hat i}\, \delta^{\hat i}_{\phantom{\hat i}i}
- v_i\, (\psi_A)^J_{\phantom{J}\hat K}\, (\psi^{\dagger A})^{\hat K}_{\phantom{\hat K} i}\, \delta^{i}_{\phantom{i}\hat i}\, (Y^\dagger_1)^{\hat i}_{\phantom{\hat i} J}
\nonumber\\&
-2\, \bar v^i\, (Y^A)^i_{\phantom{i}\hat J}\, (\psi^{\dagger 1})^{\hat J}_{\phantom{\hat J} K}\, (\psi_A)^K_{\phantom{K} \hat i}\, \delta^{\hat i}_{\phantom{\hat i}i} -2\,
v_i\, (Y^\dagger_A)^{\hat J}_{\phantom{\hat J}i}\, \delta^{i}_{\phantom{i}\hat i}\, (\psi^{\dagger A})^{\hat i}_{\phantom{\hat i}K}\, (\psi_1)^K_{\phantom{K}\hat J}
\nonumber\\& \left.
+2\, \bar v^i\, (\psi_A)^i_{\phantom{i}\hat K}\, (\psi^{\dagger 1})^{\hat K}_{\phantom{\hat K} J}\, (Y^A)^J_{\phantom{J}\hat i}\, \delta^{\hat i}_{\phantom{\hat i}i}
+2\, v_i\, (\psi_1)^J_{\phantom{J}\hat K}\, (\psi^{\dagger A})^{\hat K}_{\phantom{\hat K} i}\, \delta^{i}_{\phantom{i}\hat i}\, (Y^\dagger_A)^{\hat i}_{\phantom{\hat i}J}
\right]
\end{align}
and mass terms (where we have explicitly separated $U(N)$ and $U(M)$ indices)
\begin{equation}
-i\, M_A^{\phantom{A}B} \left\{m_i \left[(\psi^{\dagger A})^{\hat i}_{\phantom{\hat i}a}\, (\psi_B)^a_{\phantom{a}\hat i} - (\psi_B)^i_{\phantom{i}\hat a}\, (\psi^{\dagger A})^{\hat a}_{\phantom{\hat a}i} \right]
+
(m_i-m_j)\, (\psi^{\dagger A})^{\hat i}_{\phantom{\hat i}j}\, (\psi_B)^j_{\phantom{j}\hat i} \right\}
\end{equation}
They are written in terms of mass matrices $M_{AB}=\delta_{AB} - 2 \delta_{A1}$, breaking $SU(4)$ to $SU(3)\times U(1)$.
The second part of Yukawa interactions \eqref{eq:Yukawa1} does not produce mass terms, but only extra cubic vertices which are not relevant for the computation of the amplitudes we detail later. Their form can be found in the Appendix \eqref{eq:YukawaH2}.

\subsection{Scalar potential}

From the scalar potential \eqref{eq:scalar} we get modified terms whenever scalars with $A=1$ flavour index are present.
When all indices are set to 1, then the scalar potential vanishes identically, so we do not have to consider such a case.
When only a pair of indices is 1 the vertex reads (trace is understood)
\begin{align}\label{eq:scalar1}
& \frac14\, \left( Y^1 Y^{\dagger}_1 Y^{\hat A} Y^{\dagger}_{\hat A} Y^{\hat B} Y^{\dagger}_{\hat B} + Y^{\dagger}_1 Y^1 Y^{\dagger}_{\hat A} Y^{\hat A} Y^{\dagger}_{\hat B} Y^{\hat B} \right) + Y^1 Y^{\dagger}_{\hat A} Y^{\hat B} Y^{\dagger}_1 Y^{\hat A} Y^{\dagger}_{\hat B} + \nonumber\\&
- \frac12 \left[ Y^1 Y^{\dagger}_1 Y^{\hat A} Y^{\dagger}_{\hat B} Y^{\hat B} Y^{\dagger}_{\hat A}
+ Y^1 Y^{\dagger}_{\hat A} Y^{\hat A} Y^{\dagger}_1 Y^{\hat B} Y^{\dagger}_{\hat B} + 
Y^{\dagger}_1 Y^1 Y^{\dagger}_{\hat A} Y^{\hat B} Y^{\dagger}_{\hat B} Y^{\hat A} \right] 
\end{align}
from which quartic and quintic new scalar vertices are produced.
The latter will not be relevant for the one-loop amplitudes we are going to compute and we omit spelling them out (they are in any case easy to derive form the above formula). Quartic vertices are instead relevant since they could be used to construct potential triangle and fish diagrams, though wise choices of the amplitude could avoid these contributions. Nevertheless they should certainly be taken into account when computing the scalar fields self-energy. They read
\begin{align}\label{eq:quarticscalar}
&  \frac14\, |v_i|^2 \left[ (Y^{\hat A})^i_{\phantom{i} \hat J}\, (Y^{\dagger}_{\hat A})^{\hat J}_{\phantom{\hat J} K} \,(Y^{\hat B})^K_{\phantom{K} \hat L}\, (Y^{\dagger}_{\hat B})^{\hat L}_{\phantom{\hat L} i} 
+ (Y^{\dagger}_{\hat A})^{\hat i}_{\phantom{\hat i}J}\, (Y^{\hat A})^J_{\phantom{J}\hat K}\, (Y^{\dagger}_{\hat B})^{\hat K}_{\phantom{\hat K} L}\, (Y^{\hat B})^L_{\phantom{L}\hat i} + 
\right. \nonumber\\& \left. ~~~~~~~~
-2\, (Y^{\hat A})^i_{\phantom{i}\hat J}\, (Y^{\dagger}_{\hat B})^{\hat J}_{\phantom{\hat J} K}\, (Y^{\hat B})^K_{\phantom{K}\hat L}\, (Y^{\dagger}_{\hat A})^{\hat L}_{\phantom{L} i}
-2\, (Y^{\dagger}_{\hat A})^{\hat i}_{\phantom{\hat i}K}\, (Y^{\hat B})^K_{\phantom{K}\hat L}\, (Y^{\dagger}_{\hat B})^{\hat L}_{\phantom{\hat L} J}\, (Y^{\hat A})^J_{\phantom{J}\hat i}
\right] + \nonumber\\&
- \frac12\, v_i\, \bar v^j \left( (Y^{\dagger}_{\hat A})^{\hat i}_{\phantom{\hat i}K}\, (Y^{\hat A})^K_{\phantom{K}\hat j}\, \delta^{\hat j}_{\phantom{\hat j}j}\, (Y^{\hat B})^{j}_{\phantom{j}\hat J}\, (Y^{\dagger}_{\hat B})^{\hat J}_{\phantom{\hat J} i}\, \delta^{i}_{\phantom{i}\hat i} -
2\, (Y^{\dagger}_{\hat A})^{\hat i}_{\phantom{\hat i}J}\, (Y^{\hat B})^J_{\phantom{J}\hat j}\, \delta^{\hat j}_{\phantom{\hat j}j}\, (Y^{\hat A})^j_{\phantom{j}\hat K}\, (Y^{\dagger}_{\hat B})^{\hat K}_{\phantom{\hat K}i}\, \delta^i_{\phantom{i}\hat i}\right) 
\end{align}
Finally, additional contributions to the Lagrangian emerge from the part of the scalar potential with two pairs of indices equal to 1. 
This generates mass terms and cubic, quartic and quintic (which again won't play any role in this paper) interaction vertices.
The vertex reads (trace is understood)
\begin{equation}\label{eq:scalar2}
-\frac14 \left( Y^1 Y^{\dagger}_1 Y^{[1} Y^{\dagger}_1 Y^{A]} Y^{\dagger}_A + 
Y^{\dagger}_1 Y^{1} Y^{\dagger}_{[1} Y^{1} Y^{\dagger}_{A]} Y^A
 \right)
\end{equation}
Among the new pieces coming from plugging the vacuum expectation value \eqref{eq:Higgsing} of $Y^1$ are the mass terms
\begin{equation}
-\left[m_i^2 \left( (Y^{\hat A})^i_{\phantom{i}\hat I}\, (Y^{\dagger}_{\hat A})^{\hat I}_{\phantom{\hat I} i} + (Y^{\dagger}_{\hat A})^{\hat i}_{\phantom{\hat i}I}\, (Y^{\hat A})^I_{\phantom{I}\hat i} \right) - 2\, m_i\, m_j\, (Y^{\hat A})^i_{\phantom{i}\hat j}\, (Y^\dagger_{\hat A})^{\hat j}_{\phantom{\hat j}i} \right]
\end{equation}
From this mass formula one sees that diagonal fields stay massless, whereas off diagonal ones are all massive.
In particular heavy $(Y^{\hat A})^a_{\phantom{a}\hat i}$ fields have mass $m_i^2$, whereas light fields $(Y^{\hat A})^i_{\phantom{i}j}$  have mass $(m_i-m_j)^2$.
Then there are new cubic vertices
\begin{align}\label{eq:cubic-scalar}
& -\frac14\, |v_i|^2\, \left[ 
v_i\, \left( (Y^{\dagger}_{1})^{\hat i}_{\phantom{\hat i} K}\, (Y^{\hat A})^K_{\phantom{K}\hat J}\, (Y^\dagger_{\hat A})^{\hat J}_{\phantom{\hat J}i}\, \delta^i_{\phantom{i}\hat i} + (Y^{\dagger}_1)^{\hat K}_{\phantom{\hat K}i}\, \delta^{i}_{\phantom{i}\hat i} (Y^\dagger_{\hat A})^{\hat i}_{\phantom{\hat i}J}\, (Y^{\hat A})^J_{\phantom{J}\hat K} \right) + \right.\nonumber\\&
~~~~~ + \bar v^i\,  \left( (Y^1)^{i}_{\phantom{i}\hat K}\, (Y^{\dagger}_{\hat A})^{\hat K}_{\phantom{\hat K}J}\, (Y^{\hat A})^{J}_{\phantom{J}i} + (Y^1)^K_{\phantom{K}\hat i}\, \delta^{\hat i}_{\phantom{\hat i}i}\, (Y^{\hat A})^i_{\phantom{i}\hat J}\, (Y^\dagger_{\hat A})^{\hat J}_{\phantom{\hat J}K} \right) +\nonumber\\&
~~~~~ + v_j\, \left(
-2\, (Y^{\hat A})^K_{\phantom{K}\hat i}\, (Y^\dagger_{\hat A})^{\hat i}_{\phantom{\hat i}j}\, \delta^{j}_{\phantom{j}\hat j}\, (Y^\dagger_{1})^{\hat j}_{\phantom{\hat j} K} + (Y^{\hat A})^i_{\phantom{i} \hat K}\, (Y^\dagger_{\hat A})^{\hat K}_{\phantom{\hat K}j}\, (Y^\dagger_{1})^j_{\phantom{j}i} + 
\right.\nonumber\\&\left.~~~~~~~~
+ (Y^\dagger_{1})^{\hat i}_{\phantom{\hat i}j}\, (Y^\dagger_{\hat A})^j_{\phantom{j}K}\, (Y^{\hat A})^K_{\phantom{K}i}\, \delta^i_{\phantom{i}\hat i} - 2\, (Y^\dagger_{1})^{\hat K}_{\phantom{\hat K}j}\, \delta^j_{\phantom{j}\hat j}\, (Y^\dagger_{\hat A})^{\hat j}_{\phantom{\hat j}i}\, (Y^{\hat A})^i_{\phantom{i}\hat K}
\right) +\nonumber\\&
~~~~~ + \bar v^j\,\left(
-2\, (Y^\dagger_{\hat A})^{\hat K}_{\phantom{\hat K}i}\, (Y^{\hat A})^i_{\phantom{i}\hat j}\, \delta^{\hat j}_{\phantom{\hat j}j}\, (Y^1)^j_{\phantom{j}\hat K} + (Y^\dagger_{\hat A})^{\hat i}_{\phantom{\hat i}K}\, (Y^{\hat A})^K_{\phantom{K}\hat j}\, (Y^{1})^{\hat j}_{\phantom{\hat j}i} \delta^i_{\phantom{i}\hat i} + 
\right.\nonumber\\&\left.\left.~~~~~~~~
+ (Y^1)^i_{\phantom{i}\hat j}\, \delta^{\hat j}_{\phantom{\hat j}j}\, (Y^{\hat A})^j_{\phantom{j}\hat K}\, (Y^\dagger_{\hat A})^{\hat K}_{\phantom{\hat K}i} -2\, (Y^1)^K_{\phantom{K}\hat j}\, \delta^{\hat j}_{\phantom{\hat j}j}\, (Y^{\hat A})^j_{\phantom{j}\hat i}\, (Y^\dagger_{\hat A})^{\hat i}_{\phantom{\hat i}K}
\right)
\right]
\end{align}
These interactions induce new self-energy bubbles for the scalars, as well as potential triangle and box diagrams.
Quartic vertices from \eqref{eq:scalar2} can be obtained whose form is rather cumbersome and which we display in full length in \eqref{eq:quarticscalar2}.

In our computation these interactions will only contribute to the light $Y^{\hat A}$ scalars self-energy via tadpole diagrams. Focussing on the vertices which are relevant for this two-point function (namely those having $(Y^{\hat A})^i_{\phantom{i}\hat j}$ and $(Y^\dagger_{\hat A})^{\hat k}_{\phantom{\hat k}l}$ fields) we can simplify their form
\begin{align}\label{eq:quarticY1}
-\frac14 & \left[ 
(|v_i|^2+|v_k|^2-2\,|v_j|^2)\, (Y^{\hat A})^i_{\phantom{i}\hat j}\, (Y^\dagger_{A})^{\hat j}_{\phantom{\hat j}k}\, (Y^1)^k_{\phantom{k}\hat a}\, (Y^\dagger_{1})^{\hat a}_{\phantom{\hat a}i} +
\right.\nonumber\\&
+ (|v_j|^2+|v_k|^2-2\,|v_i|^2)\, (Y^\dagger_{A})^{\hat j}_{\phantom{\hat j}i}\, (Y^{A})^i_{\phantom{i}\hat k}\, (Y^\dagger_{1})^{\hat k}_{\phantom{\hat k}a}\, (Y^1)^a_{\phantom{a}\hat j} +
\nonumber\\&\left.
+ v_k\, \bar v^i\, (Y^{A})^i_{\phantom{i}\hat j}\, (Y^\dagger_{A})^{\hat j}_{\phantom{\hat j}k}\, \delta^k_{\phantom{k}\hat k} (Y^\dagger_{1})^{\hat k}_{\phantom{\hat k} a}\, (Y^1)^a_{\phantom{a}\hat i}\, \delta^{\hat i}_{\phantom{\hat i}i} +
\bar v^k\, v_i\, (Y^{A})^i_{\phantom{i}\hat j}\, (Y^\dagger_{A})^{\hat j}_{\phantom{\hat j}k}\, \delta^k_{\phantom{k}\hat k}\, (Y^\dagger_{1})^{\hat k}_{\phantom{\hat k}a}\, (Y^1)^a_{\phantom{a}\hat i}\, \delta^{\hat i}_{\phantom{\hat i}i} 
 \right]
\end{align}

\subsection{Relevant propagators}

From the kinetic terms and masses described in the previous section we can extract the propagators which are needed when computing the amplitudes below.
The heavy fields running in loops have the following massive propagators
\begin{align}\label{eq:propagators}
& \left\langle (A^{\mu})^i_{\phantom{i}a}(p) (A^{\nu})^b_{\phantom{b}j}(-p) \right\rangle = \frac12\, \frac{\delta^i_{\phantom{i}j}\, \delta^b_{\phantom{b}a}}{p^2+m_i^2} \left[ -\varepsilon^{\mu\rho\nu} p_{\rho} -i\, m_i\, \eta^{\mu\nu} + i\, \frac{p^{\mu}p^{\nu}}{p^2+\xi\, m_i} \left( -\xi+m_i \right) \right]
\nonumber\\&
\left\langle (\hat A^{\mu})^{\hat i}_{\phantom{\hat i}\hat a}(p) (\hat A^{\nu})^{\hat b}_{\phantom{\hat b}\hat j}(-p) \right\rangle = \frac12\, \frac{\delta^{\hat i}_{\phantom{\hat i}\hat j}\, \delta^{\hat b}_{\phantom{\hat b}\hat a}}{p^2+m_i^2} \left[ +\varepsilon^{\mu\rho\nu} p_{\rho} - i\, m_i\, \eta^{\mu\nu} + i\, \frac{p^{\mu}p^{\nu}}{p^2-\hat \xi\, m_i} \left( \hat \xi+m_i \right) \right]
\nonumber\\&
\left\langle (Y^{\hat A})^{i}_{\phantom{i}\hat a}(p) (Y^{\dagger}_{\hat B})^{\hat b}_{\phantom{\hat b}j}(-p) \right\rangle = -i\, \frac{\delta^{\hat A}_{\hat B}\, \delta^i_{\phantom{i}j}\, \delta^{\hat b}_{\phantom{\hat b}\hat a}}{p^2 + m_{i}^2}
\qquad\,\, \left\langle (Y^{\hat A})^a_{\phantom{a}\hat i}(p) (Y^{\dagger}_{\hat B})^{\hat j}_{\phantom{\hat j}b}(-p) \right\rangle = -i\, \frac{\delta^{\hat A}_{\hat B}\, \delta^{\hat j}_{\phantom{\hat j}\hat i}\, \delta^{a}_{\phantom{a}b}}{p^2 + m_{i}^2}
\nonumber\\&
\left\langle (Y^{1})^{i}_{\phantom{i}\hat a}(p) (Y^{\dagger}_{1})^{\hat b}_{\phantom{\hat b}j}(-p) \right\rangle = -i\, \frac{\delta^{\hat A}_{\hat B}\, \delta^i_{\phantom{i}j}\, \delta^{\hat b}_{\phantom{\hat b}\hat a}}{p^2 - \hat \xi\, m_{i}}
\qquad
\left\langle (Y^{1})^{a}_{\phantom{a}\hat i}(p) (Y^{\dagger}_{1})^{\hat j}_{\phantom{\hat j} b}(-p) \right\rangle = -i\, \frac{\delta^{\hat A}_{\hat B}\, \delta^{\hat j}_{\phantom{\hat j}\hat i}\, \delta^{a}_{\phantom{a}b}}{p^2 + \xi\, m_{i}}
\nonumber\\&
\left\langle (\psi_{A})^i_{\phantom{i}\hat a}(p) (\psi^{\dagger\, B})^{\hat b}_{\phantom{\hat b}j}(-p) \right\rangle = i\, \frac{\slashed p\, \delta_{A}^{\phantom{A}B} - i\, m_i\, M_{A}^{\phantom{A}B}}{p^2+m_i^2}\, \delta^i_{\phantom{i}j}\, \delta^{\hat b}_{\phantom{\hat b}\hat a} 
\nonumber\\&
\left\langle (\psi_{A})^a_{\phantom{a}\hat i}(p) (\psi^{\dagger\, B})^{\hat j}_{\phantom{\hat j}b}(-p) \right\rangle = i\, \frac{\slashed p\, \delta_{A}^{\phantom{A}B} + i\, m_i\, M_{A}^{\phantom{A}B}}{p^2+m_i^2}\, \delta^{\hat j}_{\phantom{\hat j}\hat i}\, \delta^{a}_{\phantom{a}b}
\end{align}
From the form of the gauge propagators we see that in general a convenient gauge choice could be $\xi=-\hat \xi=0$. This would produce an unphysical pole at $k^2=0$ in the gauge field propagator, which gets eventually cancelled against the contribution from the exchange of a (massless in this gauge) $Y^1$ field when computing gauge invariant quantities.
When using Higgsing as a means of regularizing amplitudes one sets all masses to be equal:  $m_i=m$. In this special case it should be convenient to perform the gauge choice $\xi=-\hat \xi=m$, which effectively eliminates the last term of the propagator, simplifying calculations.
On the contrary, our computation of amplitudes is sufficiently simple that we do not need to make any particular gauge choice, rather we can use cancellation of the gauge dependent parts as a consistency check of our results.
In particular, the $\xi$ dependent piece of the gluon propagators can be exposed by partial fractioning the last terms in the first two lines of \eqref{eq:propagators}.

We also spell out the propagator for light external $SU(3)$ scalar fields 
\begin{equation}\label{eq:propagatorYA}
\left\langle (Y^{\hat A})^{i}_{\phantom{i}\hat j}(p) (Y^{\dagger}_{\hat B})^{\hat k}_{\phantom{\hat k}l}(-p) \right\rangle = -i\, \frac{\delta^{\hat A}_{\hat B}\, \delta^i_{\phantom{i}l}\, \delta^{\hat k}_{\phantom{\hat k}\hat j}}{p^2 + (m_{i}-m_{j})^2}
\end{equation}
Its one-loop correction also enters the computation of scalar amplitudes through the LSZ formula.
All other propagators could be derived collecting quadratic terms from the Lagrangian above.
The propagator for external light fermions is not needed in our computation, as we explain below.

\section{Symmetry properties of the simplest one-loop amplitudes}\label{sec:amplitude}

\subsection{Dual conformally invariant massive integrals}

Before starting the computation of amplitudes we would like to get an idea of what kind of integrals we should expect to arise if they were indeed invariant under extended dual conformal symmetry.
Following \cite{Alday:2009zm} the most direct way to obtain such integrals consists in considering those emerging in the massless case written in terms of dual variables. These are invariant under ordinary $d$-dimensional inversions. Then one generalizes the squared invariants of $d$-dimensional dual variables to $(d+1)$-dimensional ones, according to \eqref{eq:extension} and inserts a $\delta$-function in the measure of integration, enforcing it to stay $d$-dimensional. This provides naturally an integral which is invariant under the generator \eqref{eq:generator}.
For example, at six points and in three space-time dimensions, in the massless case, one-loop amplitudes can be expressed in terms of the dual conformally invariant triangle integrals
\begin{equation}\label{eq:DCImassless}
\int d^3k\, \frac{\sqrt{s_{i,i+1}\, s_{i+2,i+3}\, s_{i-2,i-1}}}{k^2\, (k-p_{i,i+1})^2\, (k+p_{i-1,i-2})^2} = \int d^3 x_0\, \frac{\sqrt{x_{i,i+2}^2\, x_{i,i-2}^2\, x_{i-2,i+2}^2}}{x_{0,i}^2\, x_{0,i+2}^2\, x_{0,i-2}^2}
\end{equation}
where $i=1,2\dots 6$ here labels the external momenta, with periodic identification.
The conventions we use for dual variables and momenta are spelled out in Appendix \ref{app:A}.
Integrals \eqref{eq:DCImassless} are naturally extended to massive ones, preserving dual conformal invariance, according to the prescription \eqref{eq:extension}. This procedure yields the massive integral
\begin{equation}\label{eq:DCIintegral}
\int [d^3 x_0]\, \frac{\sqrt{\hat x_{i,i+2}^2\, \hat x_{i,i-2}^2\, \hat x_{i-2,i+2}^2}}{\hat x_{0,i}^2\, \hat x_{0,i+2}^2\, \hat x_{0,i-2}^2}
\end{equation}
The measure of integration is $[d^3\, x_0]\equiv d^4 \hat x\, \delta(\hat x^3)$, where $\hat x^3$ refers to the extra component of dual coordinates. This allows to write the integrand completely in terms of extra-dimensional dual variables. In this sense the integrand transforms trivially under extra-dimensional inversion with respect to all dual variables, including the integration point. As in the massless case, this automatically excludes lower topologies than triangles, such as bubbles.
 
Apart from \eqref{eq:DCIintegral}, we can construct additional integrands which are invariant under \eqref{eq:generator}, by using explicit factors of masses in the numerators, namely not only appearing implicitly through $\hat x_{i,j}^2$'s. This is allowed since the integrand should be invariant under $d$-dimensional Lorentz transformations and translations, but not $(d+1)$-dimensional ones. Moreover the extra-dimensional inversion transformation \eqref{eq:inversion} implies that masses transform according to $m_i \rightarrow m_i/\hat x_i^2$, which can be used to balance conformal weights in such a way that the integrand is invariant.
Therefore we can naturally allow for a more general numerator of triangle integrals
\begin{equation}\label{eq:DCIintegral2}
\int [d^3 x_0]\, \frac{A\, \sqrt{\hat x_{i,j}^2\, \hat x_{i,k}^2\, \hat x_{j,k}^2} + B\, m_{k}\, \hat x_{i,j}^2 + C\, m_{j}\, \hat x_{i,k}^2 + D\, m_{i}\, \hat x_{j,k}^2 + E\, m_i\, m_{j}\, m_{k}}{\hat x_{0,i}^2\, \hat x_{0,j}^2\, \hat x_{0,k}^2}
\end{equation}
where the coefficients $A$, $B$, $C$, $D$ and $E$ are arbitrary, meaning that each single integral is individually dual conformally invariant in this extended sense, as is easy to ascertain looking at the conformal weights of each point.

We conclude this section with a remark on the four-point case. In that situation, in the massless case, it is not possible to construct a nonvanishing triangle integral which is invariant under dual conformal transformations. Rather, the dual conformal invariant integrand is a vector box with momenta in the numerator contracted by a $\varepsilon$ tensor.
In the massive case, one can indeed construct nonvanishing integrands which are invariant under \eqref{eq:generator} with explicit powers of masses in the numerators.
However it is less clear how to obtain an invariant integrand starting from the massless one and trying to extend it in the extra-dimensional manner of \eqref{eq:DCIintegral}. In particular in the formulation with a Levi-Civita tensor, the extra-dimensional deformation clashes with the three-dimensional nature of the tensor. Nevertheless the vector integral of this formulation can be reduced to a combination of scalar integrals, where the covariance under dual conformal transformation is obscured but still valid (see e.g. \cite{Bianchi:2013pfa} for the explicit decomposition). It would be interesting to determine such a dual conformally invariant combination of scalar integrals in the massive case also and to check whether four-point amplitudes depend on it.

\subsection{Scalar six-point amplitude}

In order to check if ABJM amplitudes on the moduli space possess extended dual conformal symmetry we perform the explicit computation of the simplest possible amplitude at lowest loop order. 
We assume the planar limit $N\gg M$ and color ordering.
Then the perturbative series organises in powers of the 't Hooft coupling $\lambda = \frac{N}{k}$, which we take small.
Curiously, the simplest amplitude to compute at one loop is not a four-point, but rather a six-point one.
In particular we focus on the totally scalar amplitude
\begin{equation}\label{eq:scalaramplitude}
{\cal A}_6\, =  \left\langle Y^{\dagger}_{\hat A}(p_1)\, Y^{\hat B}(p_2)\, Y^{\dagger}_{\hat C}(p_3)\, Y^{\hat A}(p_4)\, Y^{\dagger}_{\hat B}(p_5)\, Y^{\hat C}(p_6) \right\rangle
\end{equation}
where we use light external fields, namely those with indices $(Y^{\hat A})^{i_1}_{\phantom{i_1}\, \hat i_2}$ etc. as shown in figure \ref{fig:sixpoint}.
\FIGURE{
\centering
\includegraphics[width=0.39\textwidth]{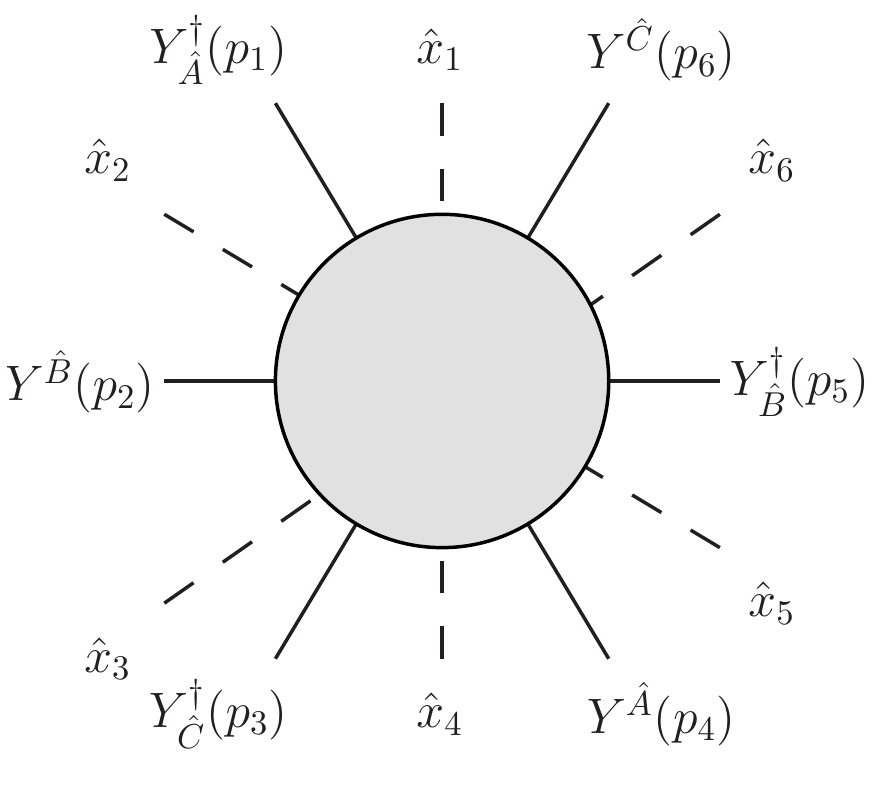}
\caption{Six scalar amplitude.}
\label{fig:sixpoint}
}
We note the particular choice of the flavour assignments which is such that at tree level the amplitude only gets contribution from a scalar potential sextic vertex.
This is true at the origin of moduli space and keeps holding at a generic point as well.
In fact it is easy to ascertain that new vertices cannot contribute to the relevant color structure of the amplitude.
This is in contrast with other amplitudes which receive additional contributions from new vertices of the Higgsed Lagrangian, already at tree level. 
The four scalar amplitude $\langle Y^\dagger_{\hat A} Y^{\hat A} Y^\dagger_{\hat B} Y^{\hat B} \rangle$ is one of those. This motivates the choice of the amplitude \eqref{eq:scalaramplitude}.
Six-point amplitudes at tree level are proportional to a factor of the coupling constant $(\frac{4\pi}{k})^2$, which we suppress in the rest of the computation.
Hence at tree level the amplitude reads 
\begin{equation}\label{eq:treeamp}
{\cal A}_6^{(0)} = \raisebox{-1.2cm}{\includegraphics[width=0.15\textwidth]{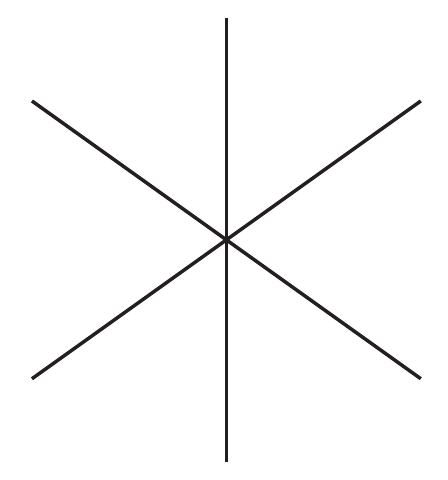}} = i
\end{equation}
Also, the particular choice of flavours dramatically constrains, in the planar limit, the number of quantum corrections this amplitude receives at one loop.
This fact was used already in the massless case to compute these amplitude by a Feynman diagram computation in \cite{Bianchi:2012cq} (although using a superspace formalism).
Also, the idea of using this kind of scalar amplitudes with a limited amount of corrections was already proposed in \cite{Alday:2009zm} in the context of ${\cal N}=4$ SYM and used in \cite{Bianchi:2012zb} (again within a superspace approach) to derive their one-loop contribution for all number of external particles.

We now compute the one-loop corrections to \eqref{eq:scalaramplitude}.
We mention (and it will also be evident from the computation that follows) that this correction vanishes identically at the origin of moduli space. This suggests that in case of dual conformal invariance, the amplitude should be expressible by integrals of the form \eqref{eq:DCIintegral2} with $A=0$, where the explicit presence of masses in the numerators guarantees that the vanishing result at the origin of the moduli space is easily recovered.

At one loop, starting with amputated graphs, there are again the same triangle diagrams as for the massless case (though with massive propagators this time), plus additional fish diagrams
\begin{align}\label{eq:diagrams}
{\cal A}_6^{(1)} \big|_{amputated} &= \sum_{i\, {\rm even}} \left( \raisebox{-1.1cm}{\includegraphics[scale=0.5]{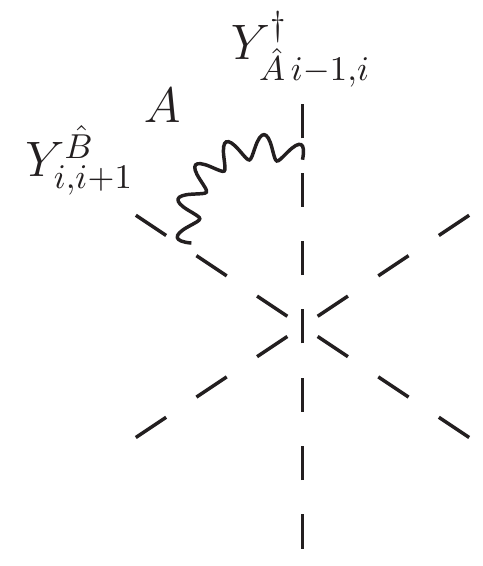}} + \raisebox{-1.1cm}{\includegraphics[scale=0.5]{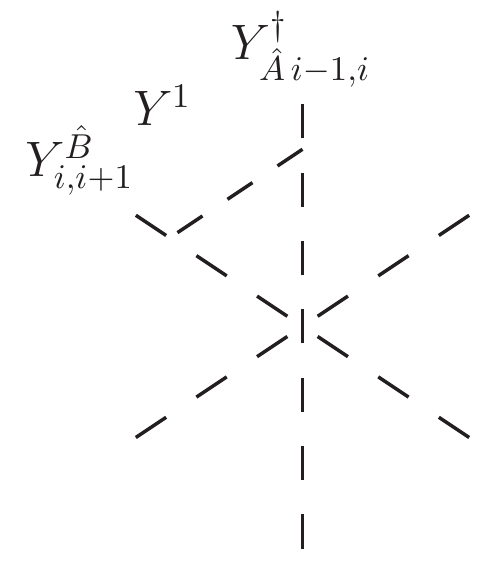}} + \raisebox{-0.6cm}{\includegraphics[scale=0.5]{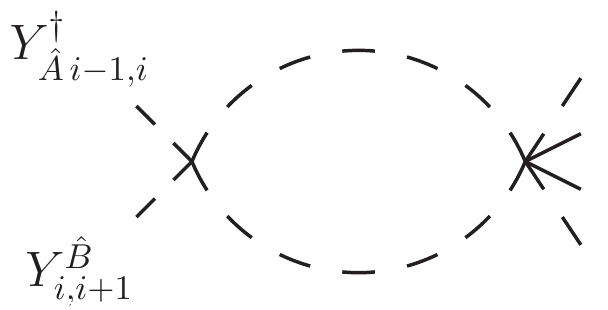}} \right) +
\nonumber\\&
+ \sum_{i\, {\rm odd}} \left( \raisebox{-1.1cm}{\includegraphics[scale=0.5]{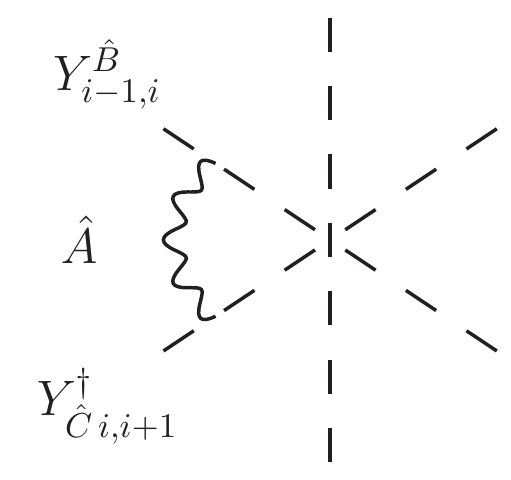}} + \raisebox{-1.05cm}{\includegraphics[scale=0.5]{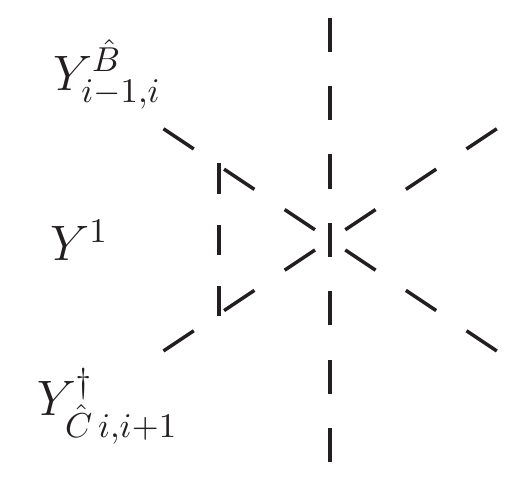}} + \raisebox{-0.65cm}{\includegraphics[scale=0.5]{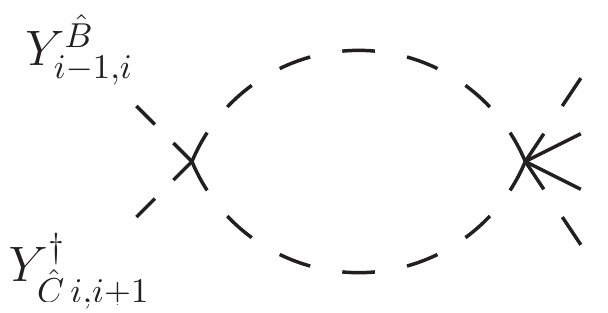}} \right) +
\nonumber\\&
+ \raisebox{-1.1cm}{\includegraphics[scale=0.5]{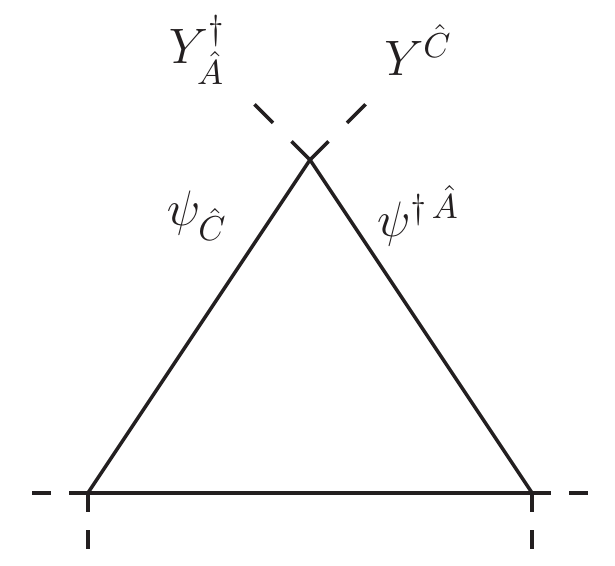}} + \raisebox{-1.1cm}{\includegraphics[scale=0.5]{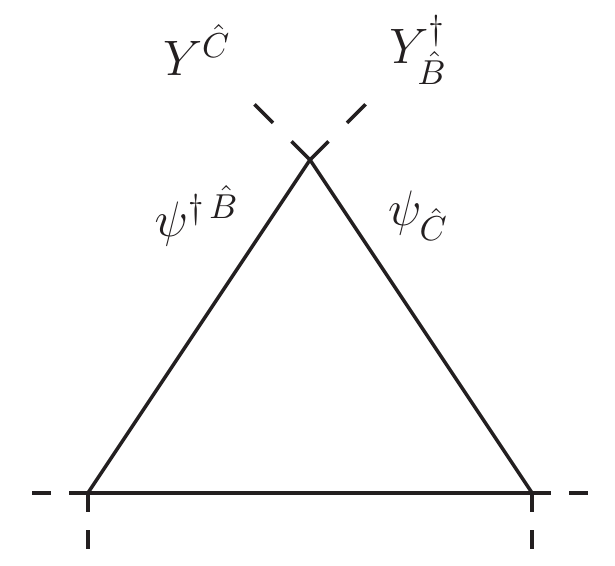}}
\end{align}
We suppress a factor $\lambda$ relative to the tree level case in the following intermediate steps.
The $i$th gauge vector $A$ exchange evaluates ($i$ is the even index of the mass corresponding to the vector boson, other masses are listed in counterclockwise order)
\begin{align}
\raisebox{-1.1cm}{\includegraphics[width=2.5cm]{gaugetriangleA}}
= \frac{i}{2}\, \frac{(k - 2p_{i})_{\mu} (k + 2p_{i-1})_{\nu}}{k^2+m_i^2}\, \left[ \varepsilon_{\mu\rho\nu} k^\rho -i\, m_i\, \eta_{\mu\nu} + i \frac{k_{\mu}k_{\nu}}{(k^2+m_i\xi)}\, (-\xi + m_i) \right] 
\end{align}
which after some algebra (in particular the part involving the Levi-Civita tensor, which is the only one contributing in the massless case, vanishes after Passarino-Veltman reduction of the vector triangle integral) gives
\begin{align}\label{eq:triangleA}
\raisebox{-1.1cm}{\includegraphics[width=2.5cm]{gaugetriangleA}}
=\frac{1}{2} & \left[
2\, (m_{i}-m_{i-1})\, B(p_{i-1}^2,m_{i-1}^2,m_{i}^2)
+ 2\, (m_{i}-m_{i+1})\, B(p_{i}^2,m_{i+1}^2,m_{i}^2) + 
\right.\nonumber\\&\left. 
- m_i\, B(p_{i,i-1}^2\,m^2_{i-1},m^2_{i+1}) + \frac{1}{m_i}\, I(m_i^2)
- 2\, \int [d^3\, x_0]\, \frac{m_i\, \hat x_{i-1,i+1}^2 }{x_{0,i}^2\, x_{0,i-1}^2\, x_{0,i+1}^2} \right]
\end{align}
where we have used $\hat x_{i-1,i+1}^2 = (p_i+p_{i+1})^2 + (m_{i-1}-m_{i+1})^2$ which naturally appears as the numerator of the triangle. The $B$ and $I$ integrals are bubbles and tadpoles with self-explanatory notation reviewed in \eqref{eq:intshort}.
There is an additional gauge dependent part, reading
\begin{align}\label{eq:triangleAgauge}
\left.\raisebox{-1.1cm}{\includegraphics[width=2.5cm]{gaugetriangleA}}\right|_{gauge\,\, dep}
 & \hspace{-1.cm} = -\frac{1}{2\, m_i}\, \Big[ I(\xi\, m_i) + m_i\,(m_i-2\,m_{i-1})\, B(p_{i-1}^2,m_{i-1}^2,\xi\, m_i) +
\nonumber\\& ~~~~~~~~
+ m_i\,(m_i-2\,m_{i+1})\, B(p_{i}^2,m_{i+1}^2,\xi\, m_i) \Big] + 
\nonumber\\& \hspace{-2.cm}
- \frac12\, \int d^3k\, \frac{m_i\, (m_i-2\,m_{i-1})(m_i-2\,m_{i+1})}{(k^2+\xi m_i)[(k-p_{i})^2+m_{i+1}^2][(k+p_{i-1})^2+m_{i-1}^2]} \qquad i\,\,{\rm even} 
\end{align}
The last term, which is a triangle, can be seen to be exactly cancelled by the same diagram where the gluon is replaced by a $Y^1$ Goldstone boson and the new cubic vertices from the scalar potential are used
\begin{equation}
\raisebox{-1.1cm}{\includegraphics[width=2.5cm]{scalartriangleY}} = \frac{1}{2}\, \int d^3k\, \frac{m_i\, (m_i-2\,m_{i-1})(m_i-2\,m_{i+1})}{(k^2+\xi m_i)[(k-p_{i})^2+m_{i+1}^2][(k+p_{i-1})^2+m_{i-1}^2]} \qquad i\,\,{\rm even} 
\end{equation}
When $i$ is odd a $\hat A$ gluon is exchanged between a pair of $Y$ and $Y^{\dagger}$ scalars in this order.
Since the part of the diagram proportional to the Levi-Civita tensor vanishes, it is easy to realize that the contribution from these diagrams is the same as in \eqref{eq:triangleA}, up to the gauge dependent part, which reads
\begin{align}
\left.\raisebox{-1.cm}{\includegraphics[width=2.5cm]{gaugetriangleAhat}}\right|_{gauge\,\, dep}
 & \hspace{-1.cm} = -\frac{1}{2\, m_i}\, \Big[ I(-\hat\xi\, m_i) + m_i\,(m_i-2\,m_{i-1})\, B(p_{i-1}^2,m_{i-1}^2,-\hat\xi\, m_i) + \nonumber\\& ~~~~~~~~~~~~
 + m_i\,(m_i-2\,m_{i+1})\, B(p_{i}^2,m_{i+1}^2,-\hat\xi\, m_i) \Big] + 
\nonumber\\& \hspace{-2.cm}
- \frac12\, \int d^3k\, \frac{m_i\, (m_i-2\,m_{i-1})(m_i-2\,m_{i+1})}{(k^2-\hat\xi m_i)[(k-p_{i})^2+m_{i+1}^2][(k+p_{i-1})^2+m_{i-1}^2]} \qquad i\,\,{\rm odd} 
\end{align}
Again, the triangle is cancelled by the exchange of a $Y^1$ scalar
\begin{equation}
\raisebox{-1.cm}{\includegraphics[width=2.5cm]{scalartriangleYd}} = \frac{1}{2}\, \int d^3k\, \frac{m_i\, (m_i-2\,m_{i-1})(m_i-2\,m_{i+1})}{(k^2-\hat\xi m_i)[(k-p_{i})^2+m_{i+1}^2][(k+p_{i-1})^2+m_{i-1}^2]} \qquad i\,\,{\rm odd} 
\end{equation}
Next we analyse the contribution of fermion loop triangles, which was also present in the massless case. Here this diagram yields
\begin{equation}\label{eq:fermiontriangle}
\raisebox{-0.8cm}{\includegraphics[width=2.5cm]{fermiontriangle2}}
= -i\, \Tr \left( \frac{k\cdot\gamma + i\, m_3\, \delta}{k^2+m_3^2}\, \frac{(k+p_{12})\cdot\gamma + i\, m_1\, \delta}{(k+p_{12})^2+m_1^2} \, \frac{(k-p_{34})\cdot\gamma + i\, m_5\, \delta}{(k-p_{34})^2+m_5^2} \right)
\end{equation}
The trace of three $\gamma$ matrices gives a Levi-Civita tensor and the resulting integral vanishes after performing Passarino-Veltman reduction.
The trace of single $\gamma$ matrices vanishes as well and one is left with the following contributions
\begin{align}
\raisebox{-0.8cm}{\includegraphics[width=2.5cm]{fermiontriangle2}}
= & - \int [d^3 x_0]\, \frac{m_3\, \hat x_{15}^2 + m_5\, \hat x_{13}^2 + m_1\, \hat x_{35}^2 + 8\, m_1\, m_3\, m_5}{\hat x_{01}^2\, \hat x_{03}^2\, \hat x_{05}^2} + 
\nonumber\\& \hspace{-3cm}
+ (m_3+m_5)\, B(p_{34}^2,m_3^2,m_5^2) + (m_1+m_5)\, B(p_{56}^2,m_1^2,m_5^2)+ (m_1+m_3)\, B(p_{12}^2,m_1^2,m_3^2)
\end{align}
The other fermion loop triangle diagram in \eqref{eq:diagrams} can be obtained rotating labels in the formula above by one site.
In fact each of the three Yukawa vertices involved in this contributions has a sign difference with respect to those used in \eqref{eq:fermiontriangle}, but it is compensated by another sign in ordering fermions when Wick contracting.
We pause at this point and focus on bubble and tadpole integrals obtained from the diagrams above, whose cancellation would represent the first hint at dual conformal symmetry.
We begin collecting bubbles having a sum of two external momenta $p_{i,i+1}$ inflowing
\begin{equation}\label{eq:sumbubble}
-\frac12\, \sum_{i=1}^6\, (m_i-2\,m_{i-1}-2\,m_{i+1})\, B(p_{i-1,i}^2,m_{i-1}^2,m_{i+1})
\end{equation}
where the sum comes from all possible gauge vector exchanges and fermion triangle diagrams.
There are potentially new bubble diagrams which can be constructed using the new scalar vertices \eqref{eq:quarticscalar} coming from the scalar potential \eqref{eq:scalar1}. On the contrary, it is easy to ascertain that those of \eqref{eq:quarticscalar2} cannot contribute to the color ordered amplitude \eqref{eq:scalaramplitude}.
These additional diagrams are depicted in the first two lines of \eqref{eq:diagrams}.
Such a contribution evaluates
\begin{equation}
\raisebox{-0.5cm}{\includegraphics[width=2.5cm]{scalarbubble}}
= \frac12\, (m_i-2\,m_{i-1}-2\,m_{i+1})\, B(p_{i-1,i}^2,m_{i-1}^2,m_{i+1}^2) 
\end{equation}
Their sum precisely cancels \eqref{eq:sumbubble}.
Then we are left with bubbles with a single momentum inflowing.
Those from \eqref{eq:triangleA} cancel out telescopically when summing over the various contributions, with periodic boundary conditions.
Finally there are gauge dependent bubbles depending on a single momentum and tadpoles of \eqref{eq:triangleA} and \eqref{eq:triangleAgauge}.
These contributions are of the same form as the corrections from the scalar self-energy entering the computation through the LSZ reduction formula.

\paragraph{Scalar self-energy}

We compute the 1PI diagrams contributing to the scalar two-point function.
Such a computation involves several diagrams summarized in figure \ref{fig:scalar-selfenergy} and its details are collected in the Appendix \ref{app:self-energy}.
\FIGURE{
\centering
\includegraphics[width=\textwidth]{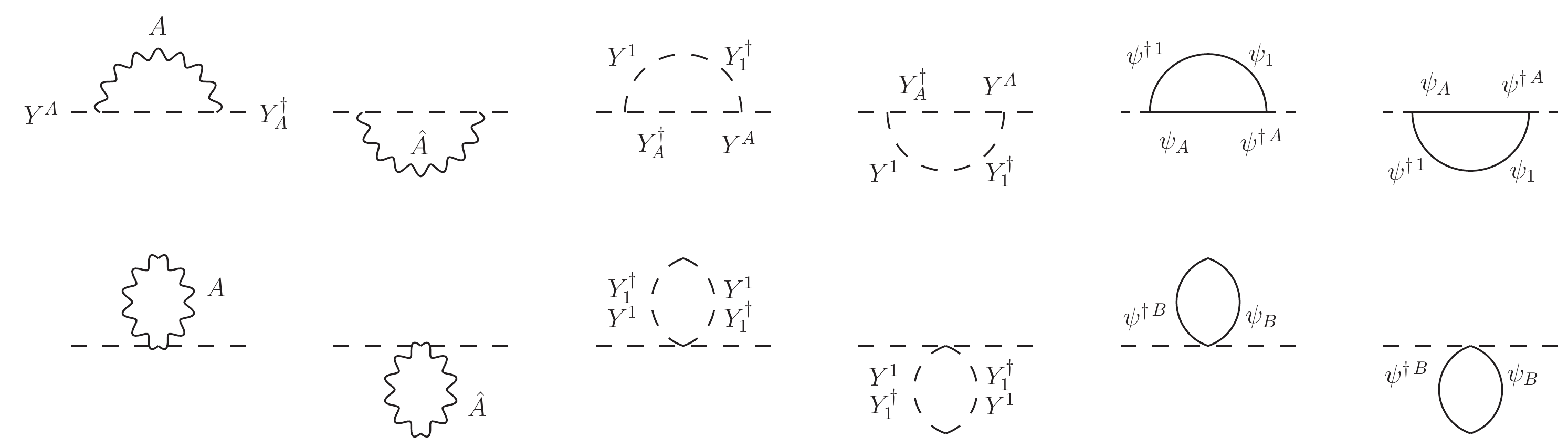}
\caption{Scalar self-energy.}
\label{fig:scalar-selfenergy}
}
Summing all these diagrams and extracting the residue at $p_i^2=-(m_{i}-m_{i+1})^2$ gives the wave-function renormalization $Z$ of the fields $Y^{\hat A}$ at one loop.
This contributes to the amplitude to the order we are considering via the LSZ formula as follows
\begin{equation}
{\cal A}_6^{(1)} = {\cal A}^{(1)}_{amputated} - \frac{i}{2}\, {\cal A}_6^{(0)}\, \sum_{Y}\, Z^{(1)}_{Y}
\end{equation}
where we have denoted by $Z^{(1)}_{Y}$ the one-loop wave function renormalization of the external scalar fields $Y$, the factor $(-i)$ comes from the scalar propagator and $1/2$ from the square root in the LSZ prescription.
In practice, using \eqref{eq:Z} and \eqref{eq:treeamp} this means that we have to add the following contribution to the amplitude (again ignoring factors of the coupling constant)
\begin{align}
& \frac{1}{2}\, \sum_{i\, even}\, \left[
- \frac{1}{2\, m_{i}}\, \Big( {\rm I}(m_i^2) - {\rm I}(\xi\, m_i) \Big) - \frac{1}{2\,m_{i+1}}\, \Big({\rm I}(m_{i+1}^2) - {\rm I}(-\hat\xi\, m_{i+1}) \Big) + \right.\nonumber\\& \left. ~~~~
+ (m_i - 2\, m_{i+1})\, {\rm B}(p_i^2,\xi\, m_i,m_{i+1}) + (m_{i+1}-2\, m_i)\, {\rm B}(p_i^2,m_i^2,-\hat\xi\, m_{i+1}) 
\right]\nonumber\\&
+ \frac{1}{2}\, \sum_{i\, odd}\, \left[
- \frac{1}{2\, m_{i+1}}\, \Big( {\rm I}(m_{i+1}^2) - {\rm I}(\xi\, m_{i+1}) \Big) - \frac{1}{2\, m_i}\, \Big({\rm I}(m_i^2) - {\rm I}(-\hat\xi\, m_i) \Big) + \right.\nonumber\\& \left. ~~~~
+ (m_{i+1} - 2\, m_i)\, {\rm B}(p_i^2,\xi\, m_{i+1},m_i) + (m_i-2\, m_{i+1})\, {\rm B}(p_i^2,m_{i+1}^2,-\hat\xi\, m_i) 
\right]\nonumber
\end{align}
which can be checked to precisely cancel the remaining bubbles and tadpoles.

\paragraph{Final result}

We are ready to state the final result for the six-scalar one-loop amplitude (neglecting coupling constants of the tree level one)
\begin{equation}
{\cal A}_6^{(1)} = - \lambda\, \int [d^3\, x_0]\, \sum_{i=1}^6\, \left(\frac{m_i\, \hat x_{i-1,i+1}^2 }{x_{0,i}^2\, x_{0,i-1}^2\, x_{0,i+1}^2} + \frac{m_i\, \hat x_{i+2,i-2}^2 + \tfrac83\, m_{i}\, m_{i+2}\, m_{i-2}}{\hat x_{0,i}^2\, \hat x_{0,i+2}^2\, \hat x_{0,i-2}^2} \right)
\end{equation}
I stress that in the final result no bubbles and tadpoles are present, which is the first condition in order for dual conformal symmetry to hold in three dimensions.
What is left are triangle integrals only, with massive denominators that resemble strikingly those of \cite{Alday:2009zm} in ${\cal N}=4$ SYM.
The numerators display an explicit presence of masses, which was expected since the amplitude vanishes in the massless limit.
Remarkably, the labels of these masses are precisely such that the integrands are invariant under the four-dimensional inversion
\begin{equation}
\hat x_i^{\mu} \rightarrow \frac{\hat x_i^{\mu}}{\hat x_i^2} \qquad\qquad i=1,\dots 6 \qquad \mu = 0,1,2
\end{equation}
which in particular entails $m_i\rightarrow \frac{m_i}{\hat x_i^2}$.
This, in addition to invariance under three-dimensional Lorentz transformations and translations gives invariance under the generator \eqref{eq:generator} in three dimensions. 
Indeed the integrals appearing in the result are precisely those of the form \eqref{eq:DCIintegral2}, pointed out before.
As a side comment, I stress the emergence of numerators with an explicit presence of mass factors which were indeed predicted in \cite{Henn:2010bk}, although they did not play a crucial role in that context.
Here we explicitly ascertain that they emerge naturally when computing amplitudes on the moduli space (of ABJM).

\subsection{Fermionic six-point amplitude}

There is another amplitude which is particularly simple to compute.
This is the totally fermionic amplitude
\begin{equation}\label{eq:fermamp}
\bar {\cal A}_6 = \left\langle\psi^{\dagger\, \hat A}(p_1)\, \psi_{\hat B}(p_2)\, \psi^{\dagger\, \hat C}(p_3)\, \psi_{\hat A}(p_4)\, \psi^{\dagger\, \hat B}(p_5)\, \psi_{\hat C}(p_6)\right\rangle
\end{equation}
of light fields with again a peculiar choice of flavour indices.
This amplitude vanishes at tree level at the origin of the moduli space, but is nonzero at one loop \cite{Bianchi:2012cq}.
At tree level there are no additional diagrams contributing to it from the Higgsed Lagrangian, therefore the amplitude is still vanishing in the massive case
\begin{equation}
\bar {\cal A}_6^{(0)} = 0
\end{equation}
A rapid analysis at one loop reveals that in the large $N$ limit there are only two contributions from the same triangle diagram with scalars running in the loop, as in the massless case
\begin{equation}
\bar {\cal A}_6^{(1)} = \raisebox{-1.1cm}{\includegraphics[scale=0.5]{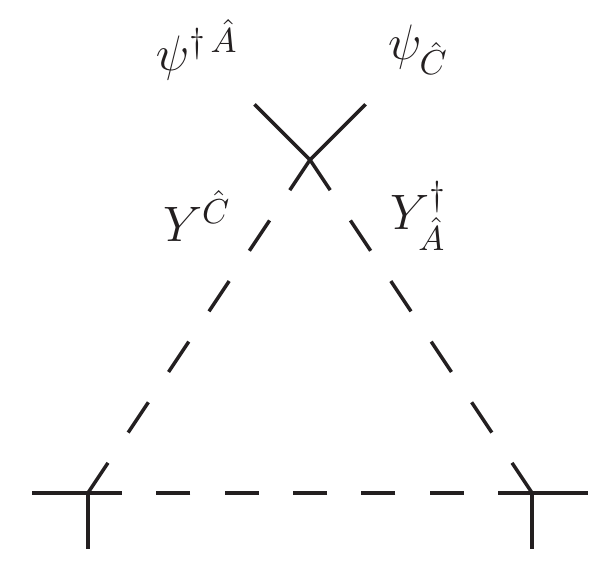}} + \raisebox{-1.1cm}{\includegraphics[scale=0.5]{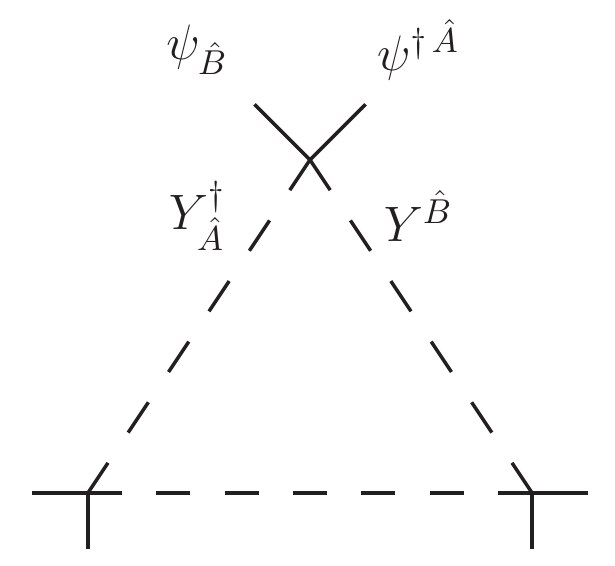}}
\end{equation}
Moreover the scalars running in the loop can only be of the $SU(3)$ sector.
These diagrams are easily evaluated and give (neglecting again coupling constants)
\begin{align}\label{eq:fermionamplitude}
\bar {\cal A}_6^{(1)} &= i \left[ \bar u(p_1)^{\alpha} u(p_2)_{\alpha} u(p_3)^{\beta} u(p_4)_{\beta} u(p_5)^{\gamma} u(p_6)_{\gamma} \int [d^3 x_0]\, \frac{1}{\hat x_{01}^2\, \hat x_{03}^2\, \hat x_{05}^2}
+ \right. \nonumber\\& \left.
~~ + \bar u(p_6)^{\alpha} u(p_1)_{\alpha} u(p_2)^{\beta} u(p_3)_{\beta} u(p_4)^{\gamma} u(p_5)_{\gamma} \int [d^3 x_0]\, \frac{1}{\hat x_{02}^2\, \hat x_{04}^2\, \hat x_{06}^2} \right]
\end{align}
The fact that the tree level amplitude vanishes also implies that this amplitude does not receive corrections from the fermion self-energy which therefore we do not need to compute.
Thus \eqref{eq:fermionamplitude} is the complete one-loop correction to \eqref{eq:fermamp}.
In the numerator there appear the polarization spinors for fermions. In the massive case they are solutions of the Dirac equation for a massive fermion
\begin{equation}
(-ip_{\mu}\gamma^{\mu} - m ) \psi =0
\end{equation}
with the standard ansatz $\psi(p) = u(p) e^{-ipx}$.
In our conventions of Appendix \ref{app:A} this solution reads \cite{Beisert:2008iq}
\begin{equation}\label{eq:polarization}
u(p) = \frac{1}{\sqrt{p_0-p_1}} \left( \begin{array}{c} p_2-im \\ p_1-p_0 \end{array}\right)
\end{equation}
which (together with its complex conjugate) can be used as a polarization spinor for in(out)coming (anti)fermions.
In the massless case the two solutions are identical $u(p_i)|_{p_i^2=0}=v(p_i)|_{p_i^2=0}=\lambda_i$ and satisfy
\begin{equation}
\langle i\, j\rangle^2 \equiv \left(\lambda_i^{\alpha}\varepsilon_{\alpha\beta} \lambda_j^\beta\right)^2 = -2\, p_i \cdot p_j
\end{equation}
Using this property, the numerator of the integrals \eqref{eq:fermionamplitude} can be rewritten in the form \eqref{eq:DCImassless} which is manifestly dual conformally invariant (and yields a constant when integrated).
In the massive case such an identification is not possible any longer and the covariant properties under inversion of the massless case are lost.
Nevertheless amplitudes with fermions would need to be embedded into a proper massive superamplitude in order to really ascertain their symmetry properties under dual (super)conformal invariance \cite{Craig:2011ws}.
Anyway it is comforting that at least the denominator of the integral has still the same form as the scalar amplitude and as expected from the ${\cal N}=4$ SYM case.
In particular the fermion amplitude \eqref{eq:fermionamplitude} is obtained from the massless integrand deforming the denominator according to the extra-dimensional prescription $x_i\rightarrow \hat x_i$ and replacing the massless polarization spinors by the massive ones.
Moreover, insisting on the suggestive extra-dimensional interpretation of amplitudes on the moduli space of \cite{Alday:2009zm}, we can regard the mass appearing in the polarization spinors \eqref{eq:polarization} as a fourth coordinate of momenta as follows
\begin{equation}
x_i \rightarrow \hat x_i = (x_i,m_i)\qquad \Rightarrow \qquad p_i \rightarrow \hat p_i = (x_{i+1}-x_i,m_{i+1}-m_i)
\end{equation}
and we define $m_{i+1}-m_i=m$ in \eqref{eq:polarization}.
Then the massive polarization spinors \eqref{eq:polarization} are morally of the same form as the four-dimensional helicity spinors $\lambda$ and $\tilde \lambda$ for massless momenta (properly identifying their components).
Therefore the massive amplitude \eqref{eq:fermionamplitude} is somehow obtainable from the massless case by translating the external kinematics to a four-dimensional one, including polarization spinors.

\section{Conclusions}

In this letter we have considered partially Higgsed ABJM theory and computed the simplest one-loop six-point amplitudes. We find that these are compatible with dual conformal invariance involving masses, which represents a strong test in favour of it to hold also away from the origin of the moduli space, as suggested in \cite{CaronHuot:2012hr}.
We have used some special six-point amplitudes since they are the easiest examples to study in terms of number and complexity of Feynman diagrams.
It would be interesting to extend this analysis to the four-point amplitude as well.
Considering the scattering of four scalars, then more Feynman diagrams are required for the evaluation of its one-loop correction.
In particular, contrary to the six-point case analysed above, box diagrams are also possible which are likely to produce scalar box integrals as in the massless case.
Then one should investigate if the combination of integrals appearing in this situation is also invariant under inversions involving masses. This is the roundabout way in which dual conformal invariance manifests itself in the four-point scalar amplitude of ABJM at the origin of the moduli space, computed with Feynman diagrams.
Hence it is likely that something similar happens when computing the same object in a nontrivial vacuum.
If this was the case, that would provide a very strong check that dual conformal invariance persists away from the origin of the moduli space of ABJM.
It would also be interesting to specialize four-point scattering to the two-mass configuration as done in \cite{Henn:2010bk,Correa:2012nk} and inspect a possible relation to the space-like cusp and Bremsstrahlung function of ABJM \cite{Forini:2012bb,Griguolo:2012iq,Lewkowycz:2013laa,Bianchi:2014laa,Correa:2014aga,Aguilera-Damia:2014bqa}.

\section*{Acknowledgements}

We thank Lorenzo Bianchi, Andi Brandhuber, Matias Leoni and Gabriele Travaglini for very useful discussions. 
This work was supported in part by the Science and Technology Facilities Council Consolidated Grant ST/L000415/1 \emph{String theory, gauge theory \& duality}.

\vfill
\newpage

\appendix

\section{Conventions and notation}\label{app:A}

We are working in three-dimensional Minkowski space with metric $\eta_{\mu\nu}=\mathrm{diag}(-1,1,1)$ and a set of $(\gamma_{\mu})_{\alpha}^{\phantom{\alpha}\beta}$ matrices satisfying
\begin{equation}
(\gamma_{\mu})_{\alpha}^{\phantom{\alpha}\beta} (\gamma_{\mu})_{\beta}^{\phantom{\beta}\gamma} = \eta_{\mu\nu}\delta_{\alpha}^{\phantom{\alpha}\gamma} + \varepsilon_{\mu\nu\rho} (\gamma^{\rho})_{\alpha}^{\phantom{\alpha}\gamma}
\end{equation}
An explicit choice could be $\gamma = (i \sigma_2,\sigma_1,\sigma_3)$.

With the aforementioned choice of $\gamma$ matrices we can determine polarization spinors for fermions as solutions of the Dirac equation
\begin{equation}
\left(
\begin{array}{cc}
-p_0-p_1 & p_2 +m\\
p_2-m & -p_0+p_1
\end{array}
\right) u(p) = 0
\end{equation}
where $\psi(p) = u(p) e^{-ipx}$.
A solution reads
\begin{equation}
u(p) = \frac{1}{\sqrt{p_0-p_1}} \left( \begin{array}{c} p_2-im \\ p_1-p_0 \end{array}\right)
\end{equation}
and the complex conjugate
\begin{equation}
v(p) = \frac{1}{\sqrt{p_0-p_1}} \left( \begin{array}{c} p_2+im \\ p_1-p_0 \end{array}\right)
\end{equation}
They satisfy the relations
\begin{equation}
u_{\alpha}v_{\beta} = - p_{\alpha\beta} + i\, m\, \varepsilon_{\alpha\beta}
\end{equation}
and
\begin{equation}
\langle i \bar i \rangle \equiv u_{\alpha}\,\varepsilon^{\alpha\beta}\,v_{\beta} = 2\,i\,m
\end{equation}

\subsection{Notation for integrals}

In the paper we use both the momentum space and dual variables description of loop integrals.
They are related through the relation
\begin{equation}
p_i \equiv x_{i+1,i} \equiv x_{i+1} - x_i
\end{equation}
Adopting the notation of \cite{Alday:2009zm} we use a hat for extra-dimensional dual coordinates
\begin{equation}
\hat x_i \equiv \left( x_i, m_i \right)
\end{equation}
with the four-dimensional component playing the role of a mass.
This massive deformation entails the on-shell condition for external momenta
\begin{equation}
p_i^2 = -m_{i+1,i}^2 = - (m_{i+1}-m_i)^2
\end{equation}
which can be alternatively stated as $\hat x_{ij}$ being light-like in four-dimensional space-time.\\
In the text we use the following shorthand notations for bubble and tadpole massive integrals
\begin{align}\label{eq:intshort}
& {\rm B}(p_i^2,m_j^2,m_k^2) \equiv \int d^3 k\, \frac{1}{\left[k^2+m^2_j\right]\left[ (k-p_i)^2+m_k^2\right]}\\
& {\rm I}(m_i^2) \equiv \int d^3 k\, \frac{1}{k^2+m^2_i}
\end{align}

\section{Remaining pieces of the Higgsed Lagrangian}

In this section we complete the Higgsed Lagrangian with the extra interaction terms which were omitted in the main text.
We start from cubic interaction involving $Y^1$ and gauge fields
\begin{align}\label{eq:cubicgaugeY1}
& 2\, v_i\, (Y^\dagger_1)^{\hat I}_{\phantom{\hat I}J}\, (A_{\mu})^J_{\phantom{J}i}\, \delta^i_{\phantom{i}\hat i}\, (\hat A^{\mu})^{\hat i}_{\phantom{\hat i}\hat I} + 2\,  \bar v^i\, (A_{\mu})^{i}_{\phantom{i}I}\, (Y^1)^I_{\phantom{I}\hat J}\, (\hat A^{\mu})^{\hat J}_{\phantom{\hat J}\hat i}\, \delta^{\hat i}_{\phantom{\hat i}i} + 2\, \bar v^i\, v_i\, (A_{\mu})^i_{\phantom{i}j}\, \delta^{j}_{\phantom{j}\hat j}\, (\hat A^{\mu})^{\hat j}_{\phantom{\hat j}\hat i}\, \delta^{\hat i}_{\phantom{\hat i}i} \nonumber\\&
- v_i\, (\hat A^{\mu})^{\hat i}_{\phantom{\hat i}\hat I}\, (\hat A^{\mu})^{\hat I}_{\phantom{\hat I}\hat J}\, (Y^\dagger_1)^{\hat J}_{\phantom{\hat J}i}\, \delta^i_{\phantom{i}\hat i} 
- \bar v_i\, (\hat A^{\mu})^{\hat I}_{\phantom{\hat I}\hat J}\, (\hat A^{\mu})^{\hat J}_{\phantom{\hat J}\hat i}\, \delta^{\hat i}_{\phantom{\hat i}i}\, (Y^1)^i_{\phantom{i}\hat I}
\nonumber\\&
- \bar v_i\, (A_{\mu})^i_{\phantom{i}I}\, (A_{\mu})^I_{\phantom{I}J}\, (Y^1)^J_{\phantom{J}\hat i}\, \delta^{\hat i}_{\phantom{\hat i}i}
- v_i\, (A_{\mu})^I_{\phantom{I}J}\, (A_{\mu})^J_{\phantom{J}i}\, \delta^i_{\phantom{i}\hat i}\, (Y^\dagger_1)^{\hat i}_{\phantom{\hat i}I} 
\end{align}
From the Yukawa interactions \eqref{eq:Yukawa1} we find the following extra cubic couplings
\begin{equation}\label{eq:YukawaH2}
-i\, v_i\, \varepsilon_{1\hat B\hat C\hat D} \, (\psi^{\dagger\, \hat B})^{\hat i}_{\phantom{\hat i}I}\, (Y^{\hat C})^I_{\phantom{I}\hat J}\, (\psi^{\dagger\, \hat D})^{\hat J}_{\phantom{\hat J}i}\, \delta^i_{\phantom{i}\hat i} + i\, \bar v^i\, \varepsilon^{1\hat B\hat C\hat D} \,
(\psi_{\hat B})^i_{\phantom{i}\hat I}\, (Y^\dagger_{\hat C})^{\hat I}_{\phantom{\hat I}J}\, (\psi_{\hat D})^J_{\phantom{J}\hat i}\, \delta^{\hat i}_{\phantom{\hat i}i}
\end{equation}
whereas mass terms do not arise, since there could not be two scalars with the same index.\\
Finally, quartic vertices from the scalar potential \eqref{eq:scalar2} read
\begin{align}\label{eq:quarticscalar2}
-\frac14 & \left[ 
|v_i|^2\, (Y^{\hat A})^I_{\phantom{I}\hat J}\, (Y^\dagger_{\hat A})^{\hat J}_{\phantom{\hat J}i}\, (Y^1)^i_{\phantom{i}\hat K}\, (Y^\dagger_{1})^{\hat K}_{\phantom{\hat K}I} -
|v_i|^2\, (Y^\dagger_{\hat A})^{\hat I}_{\phantom{\hat I}i}\, (Y^{\hat A})^i_{\phantom{i}\hat J}\, (Y^\dagger_{1})^{\hat J}_{\phantom{\hat J}K}\, (Y^1)^K_{\phantom{K}\hat I} + \right.\nonumber\\& +
v_i\, v_j\, (Y^{\hat A})^{I}_{\phantom{I}\hat J}\, (Y^\dagger_{\hat A})^{\hat J}_{\phantom{\hat J}i}\, \delta^{i}_{\phantom{i}\hat i} (Y^\dagger_{1})^{\hat i}_{\phantom{\hat i}j}\, \delta^j_{\phantom{j}\hat j}\, (Y^\dagger_{1})^{\hat j}_{\phantom{\hat j}I} -
v_i\, v_j\, (Y^\dagger_{\hat A})^{\hat i}_{\phantom{\hat i}j}\, \delta^j_{\phantom{j}\hat j} (Y^\dagger_{1})^{\hat j}_{\phantom{\hat j}I}\, (Y^{\hat A})^I_{\phantom{I}\hat J}\, (Y^\dagger_{1})^{\hat J}_{\phantom{\hat J}i}\, \delta^i_{\phantom{i}\hat i} +  \nonumber\\& +
v_i\, \bar v^j\, (Y^\dagger_{\hat A})^{\hat I}_{\phantom{\hat I}i}\, \delta^i_{\phantom{i}\hat i}\, (Y^\dagger_{1})^{\hat i}_{\phantom{\hat i}J}\, (Y^1)^J_{\phantom{J}\hat j}\, \delta^{\hat j}_{\phantom{\hat j}j}\, (Y^{\hat A})^j_{\phantom{j}\hat I} -
v_i \bar v_j\, (Y^\dagger_{\hat A})^{\hat I}_{\phantom{\hat I}i}\, \delta^i_{\phantom{i}\hat i}\, (Y^\dagger_{1})^{\hat i}_{\phantom{\hat i}J}\, (Y^{\hat A})^J_{\phantom{J}\hat j}\, \delta^{\hat j}_{\phantom{\hat j}j}\, (Y^1)^j_{\phantom{j}\hat I} + \nonumber\\& +
|v_i|^2\, (Y^{\hat A})^I_{\phantom{I}\hat J}\, (Y^\dagger_{\hat A})^{\hat J}_{\phantom{\hat J}K}\, (Y^1)^K_{\phantom{K}\hat i}\, (Y^\dagger_{1})^{\hat i}_{\phantom{\hat i}I} -
\bar v^i\, v^j\, (Y^\dagger_{A})^{\hat i}_{\phantom{\hat i}I}\, (Y^1)^I_{\phantom{I}\hat j}\, \delta^{\hat j}_{\phantom{\hat j}j}\, (Y^A)^j_{\phantom{j}\hat J}\, (Y^\dagger_{1})^{\hat J}_{\phantom{\hat J}i}\, \delta^i_{\phantom{i}\hat i} +  \nonumber\\& +
\bar v^i\, \bar v^j\, (Y^{\hat A})^i_{\phantom{i}\hat I}\, (Y^\dagger_{\hat A})^{\hat I}_{\phantom{\hat I}J}\, (Y^1)^J_{\phantom{J}\hat j}\, \delta^{\hat j}_{\phantom{\hat j}j}\, (Y^1)^j_{\phantom{j}\hat i}\, \delta^{\hat i}_{\phantom{\hat i}i} -
|v_i|^2\, (Y^\dagger_{\hat A})^{\hat i}_{\phantom{\hat i}I}\, (Y^1)^I_{\phantom{I}\hat J}\, (Y^\dagger_{1})^{\hat J}_{\phantom{\hat J}K}\, (Y^{\hat A})^K_{\phantom{K}\hat i} +  \nonumber\\& + \left.
|v_i|^2\, (Y^{\hat A})^i_{\phantom{i}\hat I}\, (Y^\dagger_{\hat A})^{\hat I}_{\phantom{\hat I}J}\, (Y^1)^J_{\phantom{J}\hat K}\, (Y^\dagger_{1})^{\hat K}_{\phantom{\hat K}i} -
\bar v^i\, \bar v^j\, (Y^\dagger_{\hat A})^{\hat I}_{\phantom{\hat I}J}\, (Y^1)^J_{\phantom{J}\hat i}\, \delta^{\hat i}_{\phantom{\hat i}i}\, (Y^{\hat A})^i_{\phantom{i}\hat j}\, \delta^{\hat j}_{\phantom{\hat j}j} (Y^1)^j_{\phantom{j}\hat I}
+ {\rm h.c.}
 \right]
\end{align}

\section{One-loop scalar self-energy}\label{app:self-energy}

In this appendix we provide details of the computation of the one-loop corrections to the two-point functions of the $SU(3)$ scalar fields $Y^{\hat A}$
\begin{equation}
\langle (Y^{\hat A})^i_{\phantom{i}\hat j} (Y_{\hat B}^\dagger)^{\hat k}_{\phantom{\hat k}l} \rangle
\end{equation}
There are nonvanishing contributions from both bubble and tadpole diagrams.
In the massless case this correction evaluates to zero, therefore the scalar self-energy here originates entirely from Higgsing the theory, from both extending propagators to massive ones and new vertices.
The relevant diagrams were pictured in figure \ref{fig:scalar-selfenergy}.
All corrections are proportional to a trivial common factor $\delta^{\hat A}_{\hat B} \, \delta^{i}_{\phantom{i}l}\, \delta^{\hat k}_{\phantom{k}\hat j}\, \frac{N}{k}$ which we strip off the following contributions.
Starting with bubble diagrams we obtain:
\begin{align}
\raisebox{0.mm}{\includegraphics[width=3cm]{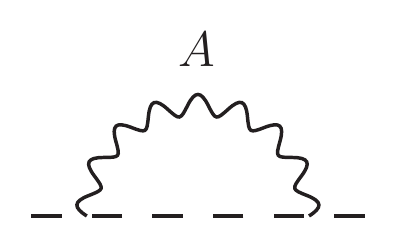}} &= \frac12\, \bigg\{ m_i\, \left[ 2\, {\rm I}(m_i^2) - {\rm I}(m_j^2) + \left( p^2 + m_i^2 - 2\, m_j^2 \right)\, {\rm B}(p^2,m_i^2,m_j^2) \right] + \nonumber\\& ~~~~~~~~
+ \frac{1}{m_i}\, \left[ (p^2+m_j^2)^2\, \left( {\rm B}(p^2,m_i^2,m_j^2) - {\rm B}(p^2,\xi\, m_i,m_j^2) \right) + \right.\nonumber\\& \left. ~~~~~~~~~~~~
- (p^2+m_i^2+m_j^2)\, {\rm I}(m_i^2) + (p^2+\xi\, m_i + m_j^2)\, {\rm I}(\xi\, m_i)
\right]
 \bigg\} \\
\raisebox{-1.cm}{\includegraphics[width=3cm]{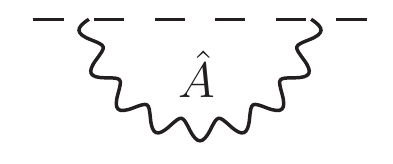}} &= \frac12\, \bigg\{ m_j\, \left[ 2\, {\rm I}(m_j^2) - {\rm I}(m_i^2) + \left( p^2 + m_j^2 - 2\, m_i^2 \right)\, {\rm B}(p^2,m_i^2,m_j^2) \right] + \nonumber\\& ~~~~~~~~
+ \frac{1}{m_j}\, \left[ (p^2+m_i^2)^2\, \left( {\rm B}(p^2,m_i^2,m_j^2) - {\rm B}(p^2,m_i^2,-\hat \xi\, m_j) \right) + \right.\nonumber\\& \left. ~~~~~~~~~~~~
- (p^2+m_i^2+m_j^2)\, {\rm I}(m_j^2) + (p^2-\hat\xi\, m_j + m_i^2)\, {\rm I}(-\hat\xi\, m_j)
\right]
 \bigg\} \\
\raisebox{-0.6 cm}{\includegraphics[width=3cm]{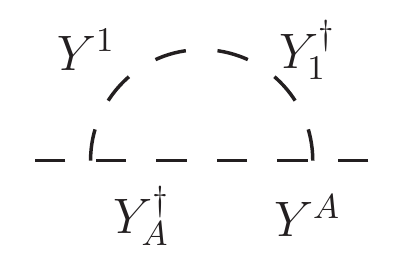}} &= \frac12\, m_i\, (m_i-2\,m_j)^2\, {\rm B}(p^2,\xi\, m_i, m_j^2) \\
\raisebox{-1. cm}{\includegraphics[width=3cm]{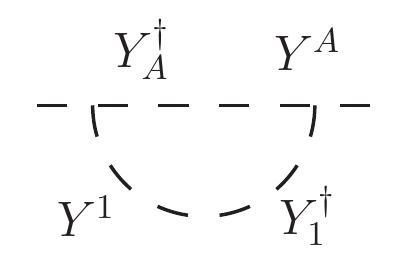}} &= \frac12\, m_j\, (m_j-2\,m_i)^2\, {\rm B}(p^2,-\hat\xi\, m_j, m_i^2) \\
\raisebox{-0.6cm}{\includegraphics[width=3cm]{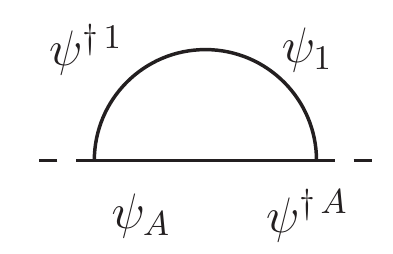}} &= 2\, m_j\, \left[ {\rm I}(m_j^2) + {\rm I}(m_i^2) - \left( p^2 + (m_i-m_j)^2 \right)\, {\rm B}(p^2,m_i^2,m_j^2) \right] \\
\raisebox{-1.cm}{\includegraphics[width=3cm]{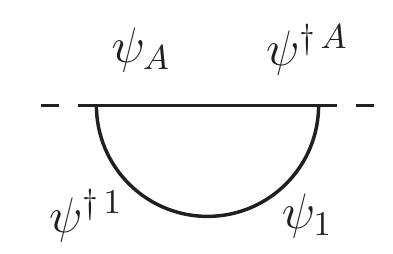}} &= 2\, m_i\, \left[ {\rm I}(m_j^2) + {\rm I}(m_i^2) - \left( p^2 + (m_i-m_j)^2 \right)\, {\rm B}(p^2,m_i^2,m_j^2) \right] \\
\end{align}
Then there are tadpole contributions reading
\begin{align}
\raisebox{-0.1 cm}{\includegraphics[width=3cm]{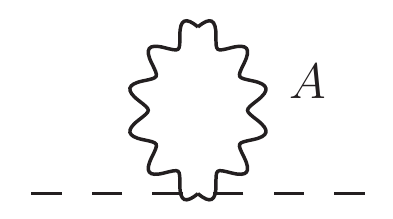}} &= -\frac12\, \left[ m_i\, (d-1)\, {\rm I}(m_i^2) + \xi\, {\rm I}(\xi\, m_i) \right]   \\
\raisebox{-1.3 cm}{\includegraphics[width=3cm]{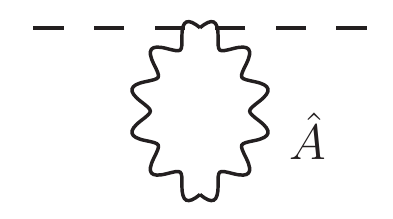}} &= -\frac12\, \left[ m_j\, (d-1)\, {\rm I}(m_j^2) - \hat \xi\, {\rm I}(-\hat \xi\, m_j) \right] \\
\raisebox{0.mm}{\includegraphics[width=3cm]{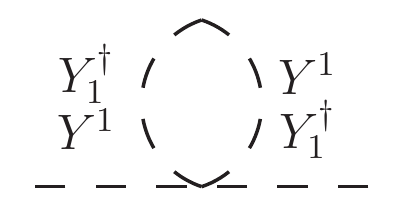}} &= -\frac12\, m_i\, {\rm I}(\xi\, m_i) - (m_i-m_j)\, {\rm I}(-\hat \xi\, m_i) \\
\raisebox{-1.3 cm}{\includegraphics[width=3cm]{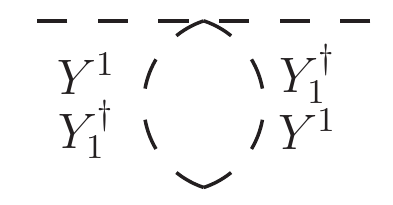}} &= -\frac12\, m_j\, {\rm I}(-\hat \xi\, m_j) - (m_j-m_i)\, {\rm I}(\xi\, m_j) \\
\raisebox{0.mm}{\includegraphics[width=3cm]{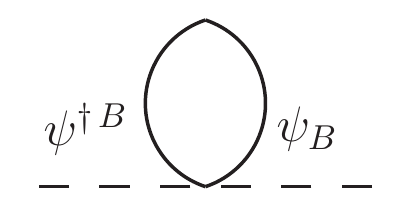}} &= -m_i\, I(m_i^2) \left( \Tr(M) -2 \right) = 0 \\
\raisebox{-1.3 cm}{\includegraphics[width=3cm]{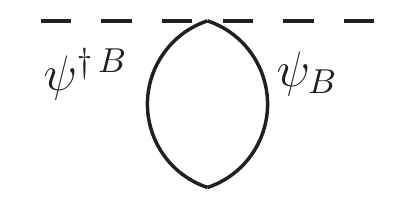}} &= -m_j\, I(m_j^2) \left( \Tr(M) -2 \right) = 0 
\end{align}
where $d$ stands for the space-time dimension ($d=3$) and the shorthand for integrals is collected in \eqref{eq:intshort}. Also note that in the totally scalar tadpoles there are two different scalar exchanges as drawn in the picture, according to the couplings \eqref{eq:quarticY1}. The contribution computed above is already the sum of these. 
The final tadpoles with fermion loops are not identically vanishing, but are so after summing over the flavours.
In addition to the above tadpole diagrams there are also those emerging from the perturbative corrections to the expectation values of the $Y^1$ scalars, which however do not contribute to the wave function renormalization of the scalars.
Summing all the diagrams and extracting the residue at the mass $(m_i-m_{i+1})^2$ we obtain the one-loop wave function renormalization $Z$ for the scalar fields, which in terms of the integrals we have introduced reads
\begin{align}\label{eq:Z}
Z_{(Y^{\hat A})^i_{\phantom{i}\hat j}}^{(1)} &= \frac{1}{2\, m_i\, m_j}\, \left[
- m_j\, \Big( {\rm I}(m_i^2) - {\rm I}(\xi\, m_i) \Big) - m_i\, \Big({\rm I}(m_j^2) - {\rm I}(-\hat\xi\, m_j) \Big) + \right.\\& \left. ~~~~
+ 2\, m_j\, (m_i - 2\, m_j)\, {\rm B}(p^2,\xi\, m_i,m_j) + 2\, m_j\, (m_j-2\, m_i)\, {\rm B}(p^2,m_i^2,-\hat\xi\, m_j) 
\right]\nonumber
\end{align}
For external conjugate scalar fields we use $Z_{(Y^\dagger_{\hat A})^{\hat i}_{\phantom{\hat i}j}}= Z_{(Y^{\hat A})^j_{\phantom{j}\hat i}}$.

\vfill
\newpage

\bibliographystyle{JHEP}

\bibliography{biblio}

\providecommand{\href}[2]{#2}\begingroup\raggedright\begin{thebibliography}{10}

\bibitem{Alday:2007hr}
L.~F. Alday and J.~M. Maldacena, {\it {Gluon scattering amplitudes at strong
  coupling}},  {\em JHEP} {\bf 0706} (2007) 064,
  [\href{http://xxx.lanl.gov/abs/0705.0303}{{\tt arXiv:0705.0303}}].

\bibitem{Drummond:2007cf}
J.~Drummond, J.~Henn, G.~Korchemsky, and E.~Sokatchev, {\it {On planar gluon
  amplitudes/Wilson loops duality}},  {\em Nucl.Phys.} {\bf B795} (2008)
  52--68, [\href{http://xxx.lanl.gov/abs/0709.2368}{{\tt arXiv:0709.2368}}].

\bibitem{Drummond:2007au}
J.~Drummond, J.~Henn, G.~Korchemsky, and E.~Sokatchev, {\it {Conformal Ward
  identities for Wilson loops and a test of the duality with gluon
  amplitudes}},  {\em Nucl.Phys.} {\bf B826} (2010) 337--364,
  [\href{http://xxx.lanl.gov/abs/0712.1223}{{\tt arXiv:0712.1223}}].

\bibitem{Brandhuber:2007yx}
A.~Brandhuber, P.~Heslop, and G.~Travaglini, {\it {MHV amplitudes in N=4 super
  Yang-Mills and Wilson loops}},  {\em Nucl.Phys.} {\bf B794} (2008) 231--243,
  [\href{http://xxx.lanl.gov/abs/0707.1153}{{\tt arXiv:0707.1153}}].

\bibitem{Drummond:2006rz}
J.~Drummond, J.~Henn, V.~Smirnov, and E.~Sokatchev, {\it {Magic identities for
  conformal four-point integrals}},  {\em JHEP} {\bf 0701} (2007) 064,
  [\href{http://xxx.lanl.gov/abs/hep-th/0607160}{{\tt hep-th/0607160}}].

\bibitem{Drummond:2007aua}
G.~Korchemsky, J.~Drummond, and E.~Sokatchev, {\it {Conformal properties of
  four-gluon planar amplitudes and Wilson loops}},  {\em Nucl.Phys.} {\bf B795}
  (2008) 385--408, [\href{http://xxx.lanl.gov/abs/0707.0243}{{\tt
  arXiv:0707.0243}}].

\bibitem{Drummond:2008vq}
J.~Drummond, J.~Henn, G.~Korchemsky, and E.~Sokatchev, {\it {Dual
  superconformal symmetry of scattering amplitudes in N=4 super-Yang-Mills
  theory}},  {\em Nucl.Phys.} {\bf B828} (2010) 317--374,
  [\href{http://xxx.lanl.gov/abs/0807.1095}{{\tt arXiv:0807.1095}}].

\bibitem{Brandhuber:2008pf}
A.~Brandhuber, P.~Heslop, and G.~Travaglini, {\it {A Note on dual
  superconformal symmetry of the N=4 super Yang-Mills S-matrix}},  {\em
  Phys.Rev.} {\bf D78} (2008) 125005,
  [\href{http://xxx.lanl.gov/abs/0807.4097}{{\tt arXiv:0807.4097}}].

\bibitem{Drummond:2009fd}
J.~M. Drummond, J.~M. Henn, and J.~Plefka, {\it {Yangian symmetry of scattering
  amplitudes in N=4 super Yang-Mills theory}},  {\em JHEP} {\bf 0905} (2009)
  046, [\href{http://xxx.lanl.gov/abs/0902.2987}{{\tt arXiv:0902.2987}}].

\bibitem{BM}
N.~Berkovits and J.~Maldacena, {\it {Fermionic T-Duality, Dual Superconformal
  Symmetry, and the Amplitude/Wilson Loop Connection}},  {\em JHEP} {\bf 0809}
  (2008) 062, [\href{http://xxx.lanl.gov/abs/0807.3196}{{\tt
  arXiv:0807.3196}}].

\bibitem{Beisert:2008iq}
N.~Beisert, R.~Ricci, A.~A. Tseytlin, and M.~Wolf, {\it {Dual Superconformal
  Symmetry from AdS(5) x S**5 Superstring Integrability}},  {\em Phys.Rev.}
  {\bf D78} (2008) 126004, [\href{http://xxx.lanl.gov/abs/0807.3228}{{\tt
  arXiv:0807.3228}}].

\bibitem{Alday:2009zm}
L.~F. Alday, J.~M. Henn, J.~Plefka, and T.~Schuster, {\it {Scattering into the
  fifth dimension of N=4 super Yang-Mills}},  {\em JHEP} {\bf 1001} (2010) 077,
  [\href{http://xxx.lanl.gov/abs/0908.0684}{{\tt arXiv:0908.0684}}].

\bibitem{Schabinger:2008ah}
R.~M. Schabinger, {\it {Scattering on the Moduli Space of N=4 Super
  Yang-Mills}},  \href{http://xxx.lanl.gov/abs/0801.1542}{{\tt
  arXiv:0801.1542}}.

\bibitem{BDS}
Z.~Bern, L.~J. Dixon, and V.~A. Smirnov, {\it {Iteration of planar amplitudes
  in maximally supersymmetric Yang-Mills theory at three loops and beyond}},
  {\em Phys.Rev.} {\bf D72} (2005) 085001,
  [\href{http://xxx.lanl.gov/abs/hep-th/0505205}{{\tt hep-th/0505205}}].

\bibitem{Henn:2010bk}
J.~M. Henn, S.~G. Naculich, H.~J. Schnitzer, and M.~Spradlin, {\it
  {Higgs-regularized three-loop four-gluon amplitude in N=4 SYM: exponentiation
  and Regge limits}},  {\em JHEP} {\bf 1004} (2010) 038,
  [\href{http://xxx.lanl.gov/abs/1001.1358}{{\tt arXiv:1001.1358}}].

\bibitem{Henn:2010ir}
J.~M. Henn, S.~G. Naculich, H.~J. Schnitzer, and M.~Spradlin, {\it {More loops
  and legs in Higgs-regulated N=4 SYM amplitudes}},  {\em JHEP} {\bf 1008}
  (2010) 002, [\href{http://xxx.lanl.gov/abs/1004.5381}{{\tt
  arXiv:1004.5381}}].

\bibitem{Correa:2012nk}
D.~Correa, J.~Henn, J.~Maldacena, and A.~Sever, {\it {The cusp anomalous
  dimension at three loops and beyond}},  {\em JHEP} {\bf 1205} (2012) 098,
  [\href{http://xxx.lanl.gov/abs/1203.1019}{{\tt arXiv:1203.1019}}].

\bibitem{Correa:2012at}
D.~Correa, J.~Henn, J.~Maldacena, and A.~Sever, {\it {An exact formula for the
  radiation of a moving quark in N=4 super Yang Mills}},  {\em JHEP} {\bf 1206}
  (2012) 048, [\href{http://xxx.lanl.gov/abs/1202.4455}{{\tt
  arXiv:1202.4455}}].

\bibitem{Caron-Huot:2014gia}
S.~Caron-Huot and J.~M. Henn, {\it {Solvable Relativistic Hydrogenlike System
  in Supersymmetric Yang-Mills Theory}},  {\em Phys.Rev.Lett.} {\bf 113}
  (2014), no.~16 161601, [\href{http://xxx.lanl.gov/abs/1408.0296}{{\tt
  arXiv:1408.0296}}].

\bibitem{ABJM}
O.~Aharony, O.~Bergman, D.~L. Jafferis, and J.~Maldacena, {\it {N=6
  superconformal Chern-Simons-matter theories, M2-branes and their gravity
  duals}},  {\em JHEP} {\bf 0810} (2008) 091,
  [\href{http://xxx.lanl.gov/abs/0806.1218}{{\tt arXiv:0806.1218}}].

\bibitem{Adam:2009kt}
I.~Adam, A.~Dekel, and Y.~Oz, {\it {On Integrable Backgrounds Self-dual under
  Fermionic T-duality}},  {\em JHEP} {\bf 0904} (2009) 120,
  [\href{http://xxx.lanl.gov/abs/0902.3805}{{\tt arXiv:0902.3805}}].

\bibitem{Adam:2010hh}
I.~Adam, A.~Dekel, and Y.~Oz, {\it {On the fermionic T-duality of the $AdS_4 x
  CP^3$ sigma-model}},  {\em JHEP} {\bf 1010} (2010) 110,
  [\href{http://xxx.lanl.gov/abs/1008.0649}{{\tt arXiv:1008.0649}}].

\bibitem{DO}
A.~Dekel and Y.~Oz, {\it {Self-Duality of Green-Schwarz Sigma-Models}},  {\em
  JHEP} {\bf 1103} (2011) 117, [\href{http://xxx.lanl.gov/abs/1101.0400}{{\tt
  arXiv:1101.0400}}].

\bibitem{Bakhmatov:2010fp}
I.~Bakhmatov, {\it {On $AdS_4 x CP^3$ T-duality}},  {\em Nucl.Phys.} {\bf B847}
  (2011) 38--53, [\href{http://xxx.lanl.gov/abs/1011.0985}{{\tt
  arXiv:1011.0985}}].

\bibitem{Bakhmatov:2011aa}
I.~Bakhmatov, E.~O~Colgain, and H.~Yavartanoo, {\it {Fermionic T-duality in the
  pp-wave limit}},  {\em JHEP} {\bf 1110} (2011) 085,
  [\href{http://xxx.lanl.gov/abs/1109.1052}{{\tt arXiv:1109.1052}}].

\bibitem{OColgain:2012si}
E.~O~Colgain, {\it {Fermionic T-duality: A snapshot review}},  {\em
  Int.J.Mod.Phys.} {\bf A27} (2012) 1230032,
  [\href{http://xxx.lanl.gov/abs/1210.5588}{{\tt arXiv:1210.5588}}].

\bibitem{BLM}
T.~Bargheer, F.~Loebbert, and C.~Meneghelli, {\it {Symmetries of Tree-level
  Scattering Amplitudes in N=6 Superconformal Chern-Simons Theory}},  {\em
  Phys.Rev.} {\bf D82} (2010) 045016,
  [\href{http://xxx.lanl.gov/abs/1003.6120}{{\tt arXiv:1003.6120}}].

\bibitem{HL2}
Y.-t. Huang and A.~E. Lipstein, {\it {Amplitudes of 3D and 6D Maximal
  Superconformal Theories in Supertwistor Space}},  {\em JHEP} {\bf 1010}
  (2010) 007, [\href{http://xxx.lanl.gov/abs/1004.4735}{{\tt
  arXiv:1004.4735}}].

\bibitem{Bianchi:2012cq}
M.~S. Bianchi, M.~Leoni, A.~Mauri, S.~Penati, and A.~Santambrogio, {\it {One
  Loop Amplitudes In ABJM}},  {\em JHEP} {\bf 1207} (2012) 029,
  [\href{http://xxx.lanl.gov/abs/1204.4407}{{\tt arXiv:1204.4407}}].

\bibitem{Bargheer:2012cp}
T.~Bargheer, N.~Beisert, F.~Loebbert, T.~McLoughlin, N.~Beisert, et~al., {\it
  {Conformal Anomaly for Amplitudes in $\mathcal{N}=6$ Superconformal
  Chern-Simons Theory}},  {\em J.Phys.} {\bf A45} (2012) 475402,
  [\href{http://xxx.lanl.gov/abs/1204.4406}{{\tt arXiv:1204.4406}}].

\bibitem{Brandhuber:2012un}
A.~Brandhuber, G.~Travaglini, and C.~Wen, {\it {A note on amplitudes in N=6
  superconformal Chern-Simons theory}},  {\em JHEP} {\bf 1207} (2012) 160,
  [\href{http://xxx.lanl.gov/abs/1205.6705}{{\tt arXiv:1205.6705}}].

\bibitem{Brandhuber:2012wy}
A.~Brandhuber, G.~Travaglini, and C.~Wen, {\it {All one-loop amplitudes in N=6
  superconformal Chern-Simons theory}},  {\em JHEP} {\bf 1210} (2012) 145,
  [\href{http://xxx.lanl.gov/abs/1207.6908}{{\tt arXiv:1207.6908}}].

\bibitem{CH}
W.-M. Chen and Y.-t. Huang, {\it {Dualities for Loop Amplitudes of N=6
  Chern-Simons Matter Theory}},  {\em JHEP} {\bf 1111} (2011) 057,
  [\href{http://xxx.lanl.gov/abs/1107.2710}{{\tt arXiv:1107.2710}}].

\bibitem{BLMPS1}
M.~S. Bianchi, M.~Leoni, A.~Mauri, S.~Penati, and A.~Santambrogio, {\it
  {Scattering Amplitudes/Wilson Loop Duality In ABJM Theory}},  {\em JHEP} {\bf
  1201} (2012) 056, [\href{http://xxx.lanl.gov/abs/1107.3139}{{\tt
  arXiv:1107.3139}}].

\bibitem{BLMPS2}
M.~S. Bianchi, M.~Leoni, A.~Mauri, S.~Penati, and A.~Santambrogio, {\it
  {Scattering in ABJ theories}},  {\em JHEP} {\bf 1112} (2011) 073,
  [\href{http://xxx.lanl.gov/abs/1110.0738}{{\tt arXiv:1110.0738}}].

\bibitem{CaronHuot:2012hr}
S.~Caron-Huot and Y.-t. Huang, {\it {The two-loop six-point amplitude in ABJM
  theory}},  {\em JHEP} {\bf 1303} (2013) 075,
  [\href{http://xxx.lanl.gov/abs/1210.4226}{{\tt arXiv:1210.4226}}].

\bibitem{Bianchi:2014iia}
M.~S. Bianchi and M.~Leoni, {\it {On the ABJM four-point amplitude at three
  loops and BDS exponentiation}},  {\em JHEP} {\bf 1411} (2014) 077,
  [\href{http://xxx.lanl.gov/abs/1403.3398}{{\tt arXiv:1403.3398}}].

\bibitem{Berenstein:2008dc}
D.~Berenstein and D.~Trancanelli, {\it {Three-dimensional N=6 SCFT's and their
  membrane dynamics}},  {\em Phys.Rev.} {\bf D78} (2008) 106009,
  [\href{http://xxx.lanl.gov/abs/0808.2503}{{\tt arXiv:0808.2503}}].

\bibitem{Lee:2010hk}
K.-M. Lee and S.~Lee, {\it {1/2-BPS Wilson Loops and Vortices in ABJM Model}},
  {\em JHEP} {\bf 1009} (2010) 004,
  [\href{http://xxx.lanl.gov/abs/1006.5589}{{\tt arXiv:1006.5589}}].

\bibitem{MOS1}
J.~Minahan, O.~Ohlsson~Sax, and C.~Sieg, {\it {Anomalous dimensions at four
  loops in N=6 superconformal Chern-Simons theories}},  {\em Nucl.Phys.} {\bf
  B846} (2011) 542--606, [\href{http://xxx.lanl.gov/abs/0912.3460}{{\tt
  arXiv:0912.3460}}].

\bibitem{Bianchi:2013pfa}
L.~Bianchi and M.~S. Bianchi, {\it {Non-planarity through unitarity in ABJM}},
  {\em Phys.Rev.} {\bf D89} (2014), no.~12 125002,
  [\href{http://xxx.lanl.gov/abs/1311.6464}{{\tt arXiv:1311.6464}}].

\bibitem{Bianchi:2012zb}
M.~S. Bianchi, M.~Leoni, and S.~Penati, {\it {Minimally helicity violating,
  maximally simple scalar amplitudes in N=4 SYM}},  {\em JHEP} {\bf 1210}
  (2012) 198, [\href{http://xxx.lanl.gov/abs/1208.0329}{{\tt
  arXiv:1208.0329}}].

\bibitem{Craig:2011ws}
N.~Craig, H.~Elvang, M.~Kiermaier, and T.~Slatyer, {\it {Massive amplitudes on
  the Coulomb branch of N=4 SYM}},  {\em JHEP} {\bf 1112} (2011) 097,
  [\href{http://xxx.lanl.gov/abs/1104.2050}{{\tt arXiv:1104.2050}}].

\bibitem{Forini:2012bb}
V.~Forini, V.~G.~M. Puletti, and O.~Ohlsson~Sax, {\it {The generalized cusp in
  $AdS_4$ x $CP^3$ and more one-loop results from semiclassical strings}},
  {\em J.Phys.} {\bf A46} (2013) 115402,
  [\href{http://xxx.lanl.gov/abs/1204.3302}{{\tt arXiv:1204.3302}}].

\bibitem{Griguolo:2012iq}
L.~Griguolo, D.~Marmiroli, G.~Martelloni, and D.~Seminara, {\it {The
  generalized cusp in ABJ(M) N = 6 Super Chern-Simons theories}},  {\em JHEP}
  {\bf 1305} (2013) 113, [\href{http://xxx.lanl.gov/abs/1208.5766}{{\tt
  arXiv:1208.5766}}].

\bibitem{Lewkowycz:2013laa}
A.~Lewkowycz and J.~Maldacena, {\it {Exact results for the entanglement entropy
  and the energy radiated by a quark}},  {\em JHEP} {\bf 1405} (2014) 025,
  [\href{http://xxx.lanl.gov/abs/1312.5682}{{\tt arXiv:1312.5682}}].

\bibitem{Bianchi:2014laa}
M.~S. Bianchi, L.~Griguolo, M.~Leoni, S.~Penati, and D.~Seminara, {\it {BPS
  Wilson loops and Bremsstrahlung function in ABJ(M): a two loop analysis}},
  {\em JHEP} {\bf 1406} (2014) 123,
  [\href{http://xxx.lanl.gov/abs/1402.4128}{{\tt arXiv:1402.4128}}].

\bibitem{Correa:2014aga}
D.~H. Correa, J.~Aguilera-Damia, and G.~A. Silva, {\it {Strings in $AdS_4
  \times \mathbb{CP}^{3}$ Wilson loops in $\mathcal N=$6 super
  Chern-Simons-matter and bremsstrahlung functions}},  {\em JHEP} {\bf 1406}
  (2014) 139, [\href{http://xxx.lanl.gov/abs/1405.1396}{{\tt
  arXiv:1405.1396}}].

\bibitem{Aguilera-Damia:2014bqa}
J.~Aguilera-Damia, D.~H. Correa, and G.~A. Silva, {\it {Semiclassical partition
  function for strings dual to Wilson loops with small cusps in ABJM}},
  \href{http://xxx.lanl.gov/abs/1412.4084}{{\tt arXiv:1412.4084}}.

\end{thebibliography}\endgroup

\end{document}